\documentclass[twocolumn]{aastex63}
\usepackage{courier}
\usepackage[T1]{fontenc}
\usepackage{mathtools}
\usepackage{changepage} 
\usepackage{url}
\usepackage{appendix}
\DeclareUnicodeCharacter{2212}{-}
\usepackage{lineno}
\usepackage{gensymb}

\title{Chandra - BAT}

\newcommand{\fix}{\textcolor{red}}

\newcommand{\swift}{\textit{Swift}}
\newcommand{\myt}{\texttt{MYTorus} }
\newcommand{\bor}{\texttt{borus02} }
\newcommand{\ct}{CT-AGNs }

\shorttitle{Chandra-BAT}
\shortauthors{Silver et al.}

\begin{document}

\title{\textit{Chandra} Follow-up Observations of \textit{Swift}-BAT-selected AGNs II}

\author{R. Silver}
\affiliation{Department of Physics and Astronomy, Clemson University,  Kinard Lab of Physics, Clemson, SC 29634, USA}

\author{N. Torres-Alb\`{a}}
\affiliation{Department of Physics and Astronomy, Clemson University,  Kinard Lab of Physics, Clemson, SC 29634, USA}

\author{X. Zhao}
\affiliation{Harvard-Smithsonian Center for Astrophysics, 60 Garden Street, Cambridge, MA 02138, USA}
\affiliation{Department of Physics and Astronomy, Clemson University,  Kinard Lab of Physics, Clemson, SC 29634, USA}

\author{S. Marchesi}
\affiliation{INAF - Osservatorio di Astrofisica e Scienza dello Spazio di Bologna, Via Piero Gobetti, 93/3, 40129, Bologna, Italy}
\affiliation{Department of Physics and Astronomy, Clemson University,  Kinard Lab of Physics, Clemson, SC 29634, USA}

\author{A. Pizzetti}
\affiliation{Department of Physics and Astronomy, Clemson University,  Kinard Lab of Physics, Clemson, SC 29634, USA}

\author{M. Ajello}
\affiliation{Department of Physics and Astronomy, Clemson University,  Kinard Lab of Physics, Clemson, SC 29634, USA}

\author{G. Cusumano}
\affiliation{INAF-Istituto di Astrofisica Spaziale e Fisica Cosmica, Via Ugo la Malfa, 153, I-90146 Palermo PA, Italy}

\author{V. La Parola}
\affiliation{INAF-Istituto di Astrofisica Spaziale e Fisica Cosmica, Via Ugo la Malfa, 153, I-90146 Palermo PA, Italy}

\author{A. Segreto}
\affiliation{INAF-Istituto di Astrofisica Spaziale e Fisica Cosmica, Via Ugo la Malfa, 153, I-90146 Palermo PA, Italy}

\author{A. Comastri}
\affiliation{INAF - Osservatorio di Astrofisica e Scienza dello Spazio di Bologna, Via Piero Gobetti, 93/3, 40129, Bologna, Italy}

\begin{abstract}
We present the combined \textit{Chandra} and \textit{Swift}-BAT spectral analysis of nine low-redshift ($z \leq 0.10$), candidate heavily obscured active galactic nuclei (AGN) selected from the \textit{Swift}-BAT 150-month catalog. \textbf{We located soft (1$-$10\,keV) X-ray counterparts to these BAT sources and joint fit their spectra with physically motivated models.} The spectral analysis in the 1$-$150\,keV energy band determined that all sources are obscured, with a line-of-sight column density N$_H \geq$ 10$^{22}$ cm$^{-2}$ at a 90\% confidence level. Four of these sources show significant obscuration with N$_H \geq$ 10$^{23}$ cm$^{-2}$ and two additional sources are candidate Compton-thick Active Galactic Nuclei (CT-AGNs) with N$_H \geq$ 10$^{24}$ cm$^{-2}$. {\bf These two sources, 2MASX J02051994−0233055 and IRAS 11058$-$1131,} are the latest addition to the previous 3 CT-AGN candidates found using our strategy for soft X-ray follow-up of BAT sources. \textbf{ In here we present the results of our methodology so far, and analyze the effectiveness of applying different selection criteria to discover CT-AGN in the local Universe. Our selection criteria has a $\sim$20\% success rate of discovering heavily obscured AGN whose CT nature is confirmed by follow-up \textit{NuSTAR} observations. This is much higher than the $\sim$5\% found in blind surveys.}\\ \\


\end{abstract}

\section{Introduction} 
Current models have concluded the Cosmic X-ray Background (CXB), the diffuse X-ray emission in the 1 to $\sim$200$-$300\,keV band, is primarily produced by accreting supermassive black holes (SMBHs), i.e., active galactic nuclei \citep[AGNs,][]{Alexander_2003, Gandhi2003, Gilli2007, Treister_2009, Ueda_2014}. While the CXB emission below 10\,keV has been almost entirely resolved \citep{Worsley2005, Hickox_2006}, at $\sim$30\,keV \citep[the peak of the CXB,][]{Ajello_2008}, only 30\% of the emission is accounted for by current observations \citep{Aird_2015, Civano_2015, Mullaney_2015, Harrison_2016}. It is expected that a considerable fraction (10-20\%) of the CXB is produced by a numerous population of heavily obscured, the so-called Compton thick (CT-) AGN, which have intrinsic obscuring hydrogen column densities (N$_H$) $\geq$ 10$^{24}$ cm$^{-2}$ \citep{Risaliti_1999, Alexander_2003, Gandhi2003, Gilli2007, Treister_2009, Ueda_2014, Ananna2019}. However, only 5$-$7\% of the hard X-ray detected low-$z$ AGNs are classified as CT-AGNs \citep[]{Comastri_2004, Della_2008, Burlon2011, Ricci_2015, Lanzuisi2018} which is much lower than those predicted by population synthesis models that aim to explain the CXB \citep[30$-$50\%; see, e.g.,][]{Gilli2007, Ueda_2014, Ananna2019}. Currently, there are only $\sim$30-35 \textit{NuSTAR}-confirmed CT-AGN in the range $z \leq 0.05$ \textbf{\citep{TorresAba2021}}\footnote{See full list at \url{https://science.clemson.edu/ctagn/ctagn/}}. The low number severely limits any science designed to study this population. Detecting new CT-AGN in X-rays is crucial to advance the field, in order to explain the shape of the CXB, and to study torus properties via complex models.\\
\indent The emission from \ct is heavily suppressed below 10\,keV due to the heavy obscuration of the dusty gas surrounding the SMBH, which makes detecting CT-AGN in X-rays difficult. Their spectra is dominated by the Compton hump at $\sim$20$-$40\,keV, originating from the intrinsic emission being reflected in the torus. 

\indent Several models developed over the past decade successfully describe the X-ray emission of AGN reprocessed by the torus material, e.g., \texttt{pexrav} \citep{Magdziarz1995}; \myt \citep{Murphy2009, Yaqoob2012}; \texttt{BNtorus} \citep{Brightman2011}; \texttt{ctorus} \citep{Liu2014}; \bor \citep{Balo_borus2018}; \texttt{UXClumpy} \citep{Buchner2019}; \texttt{XClumpy} \citep{xclumpy2019}. These models are built adopting different assumptions on the geometrical distribution of the obscuring material (e.g., homogeneous vs clumpy) or its chemical composition. The clumpy models are significant because numerous works have observed variability in the line-of-sight column density, suggesting that a patchy, non-homogeneous distribution of obscuring material is favored by observational data \citep{Risaliti2002, Bianchi2012, Torricelli2014}. However, implementing these models requires high-quality data in order to break the various degeneracies between parameters. It is thus necessary to increase our pool X-ray CT-AGN, which let the reflection component shine through thanks to the suppressed line-of-sight. Previous works have used high-quality X-ray data to successfully constrain torus parameters \citep[e.g.][]{Zhao2020}.\\
\indent Instruments such as the \textit{Swift} X-Ray Telescope (\textit{Swift}-XRT) on board the Neil Gehrels \textit{Swift} satellite \citep{Gehrels04}, \textit{Chandra}, and XMM-\textit{Newton}, 
sensitive in the $\sim$0.3$-$10\,keV energy range, can only detect the Compton hump if the source's spectrum is largely redshifted ($z > 1$). In the local universe ($z < 0.1$),
an instrument with sensitivity above 10\,keV is necessary to detect and characterize CT-AGNs \citep[e.g.,][]{Marchesi17a}. The wide-field (120$\times$90 deg$^2$) Burst Alert Telescope \citep{Barthelmy2005} continually observes the whole sky in the 15$-$200\,keV band. \textit{Swift}-BAT is thus an excellent tool to create a census of the hard X-ray
emitting sources in the local universe. The combination of \textit{Swift}-BAT and soft X-ray instruments has previously proved successful in selecting and identifying candidate \ct \citep{Burlon2011, Vasudevan_2013, Ricci_2015, Koss2016, Marchesi17a, Marchesi_2017b}. \\
\indent In this work, we perform a joint \textit{Chandra}--\textit{Swift}-BAT spectral fitting in the 1.0$-$150\,keV band of nine AGN detected by \textit{Swift}-BAT in 150 months of observations. 
The joint analysis of \textit{Chandra} and \textit{Swift}-BAT data provides an ample opportunity to constrain the column density of the obscuring material surrounding the accreting SMBHs and possibly identify new candidate CT-AGNs. The aim of this work is therefore to obtain a first estimate of the line-of-sight column density for these sources, which have never been observed before in soft X-rays. Thanks to this estimate, obtained via our quick snapshot Chandra program (PI: Marchesi, see Section \ref{sec:data_reduc}), we find promising targets to follow-up with joint \textit{NuSTAR} and \textit{XMM-Newton} observations. This program will allow to obtain new high-quality data of promising CT-AGN candidates to confirm their nature, which will add to the limited pool of high-quality CT-AGNs that can be used to constrain torus properties \citep[e.g.][]{Zhao2020} and CXB models \citep[e.g.][]{Ananna2019}. The second objective of this work is to present the results of our program so far, and discuss the efficiency of the selection criteria we have used to target new CT-AGN in the local Universe. \\
\indent In addition, \textit{Chandra's} unparalleled spatial resolution ($\sim$0.5$\arcsec$) allows us to detect X-ray emission extended out to kiloparsec scales. Recent works have discovered CT-AGN with extended emission in the 3-7\,keV band primarily aligned with the ionization cones, but also present in the orthogonal direction  \citep[the cross-cones; see e.g.,][]{Fabbiano2017, Fabbiano2018, Jones2020, Ma2020, Jones2021}. The study of the extended emission in CT-AGN has the power to
constrain the duty cycle for the AGN feedback onto the host galaxy ISM \citep{Ma2020}.

This work is organized as follows: Section \ref{sec:data_reduc} describes the selection criteria used to identify potential CT-AGN candidates and the data reduction process. Section \ref{sec:all_models} discusses the models used in our spectral fitting and Section \ref{sec:sources} describes the derived results. Section \ref{sec:ext_emi} reports the findings on the extended emission. Section \ref{sec:disc} summarizes the conclusions and future work. All errors reported are at a 90\% confidence level. Standard cosmological parameters are as follows: H$_0$ = 70 km s$^{-1}$ Mpc$^{-1}$, q$_0$ = 0.0, and $\Lambda$ = 0.73. \\

\section{Selection Criteria and Data Reduction} \label{sec:data_reduc}
Since its launch, \textit{Swift}-BAT has continuously observed the hard X-ray sky. Data from the first 150 months have been combined into a catalog containing sources detected with fluxes down to $f \sim$ 3.3$\times$10$^{-12}$ erg s$^{-1}$ cm$^{-2}$ in the 15$-$150\,keV band (Segreto et al. in preparation\footnote{The 150 month catalog can be found here: \url{https://science.clemson.edu/ctagn/bat-150-month-catalog/}}). 
Due to its high sensitivity and ability to cover the entire sky, \textit{Swift}-BAT provides an excellent tool to study the hard X-ray AGN population in the nearby Universe. \\
\indent The \textit{Swift}-BAT data are processed by the BAT\_IMAGER code \citep{Segreto2010}.
This code was developed to analyze data from coded mask instruments and can perform screening, mosaicking, and source detection. All spectra used in this work have been background subtracted and were acquired by averaging over the entire \textit{Swift}-BAT exposure. The standard BAT spectral redistribution matrix was used\footnote{\url{https://heasarc.gsfc.nasa.gov/docs/heasarc/caldb/data/swift/bat/index.html}}. \\
\indent The sample in this work is selected from the 150 month catalog, which includes a total of 724 galaxies\footnote{We note that blazars were removed and 3\% of sources from the catalog do not have a determined $z$.} in the local (z $<$ 0.10, D $\lesssim$ 400 Mpc) Universe. The first step in our three-step program is to select previously unobserved sources within these 724 to propose for quick ($\sim 10$ ks) \textit{Chandra} observations. In order to find promising CT-AGN candidates, we implement the following criteria:
\begin{itemize}
\item We select sources at high Galactic latitude ($\lvert b \rvert >$ 10$\degree$) without a ROSAT \citep{Voges1999} counterpart. The emission in the 0.1 $-$ 2.4\,keV\footnote{\url{https://heasarc.gsfc.nasa.gov/docs/rosat/ass.html}} band, in which ROSAT is sensitive, is easily suppressed in heavily obscured AGN and the lack of this counterpart already suggests an expected column density logN$_H>$ 23 \citep{Ajello2008, Koss2016}. The added requirement of high Galactic latitude ensures that the obscuration responsible for the lack of ROSAT counterpart is not coming from our own galaxy.
\item We only select Seyfert 2 galaxies (Sy2s) or sources classified as galaxies. The absence of broad lines in Sy2s implies the presence of obscuring material in our line of sight. This is significant as $\sim$95\% of Seyfert 2 galaxies are obscured with a column density logN$_H>$ 22 \citep[Figure 9,][]{Koss2017}. \textbf{Sources optically classified as galaxies are likewise likely to be potentially obscured AGN. This is because a normal galaxy cannot emit strongly enough in the $15-150$ keV range to be detected by BAT, and must therefore be an AGN. Any AGN that is optically not classified as such is likely to be obscured. We note that our nine sources have optical spectra available and lack typical AGN features, hence they are likely obscured.}
\item Finally, this sample is limited to sources in the nearby universe because CT-AGN are detected at much lower redshifts than the rest of the AGN population \citep{Burlon2011}. Approximately 90\% of the \textit{Swift}-BAT-detected CT-AGN have been discovered at $z$ $\leq$ 0.1, while unobscured or Compton-thin AGN can be found up to z$\sim$0.3 \citep{Ricci2017}.
\end{itemize}
The above selection criteria have proved capable of discovering heavily obscured AGN candidates in the past, as described in \cite{Marchesi17a}.
The last nine sources for which we obtained \textit{Chandra} observing time following the described criteria are analysed in this work, and listed in Table \ref{tab:sources}. 

\indent All sources were observed with 10\,ks by \textit{Chandra} ACIS-S as a part of the \textit{Chandra} general observing Cycles 19 and 21 (Proposal numbers 19700430 and 21700085, P.I. Marchesi\footnote{\url{https://cxc.harvard.edu/cgi-bin/propsearch/prop\_details.cgi?pid=5182}} \footnote{\url{https://cxc.harvard.edu/cgi-bin/propsearch/prop\_details.cgi?pid=5667}}). The CIAO \citep{Fruscione2006} 4.12 software was used to reduce the data following standard procedures. Source and background spectra were extracted utilizing the CIAO \texttt{specextract} tool. Source spectra were calculated with a 5$\arcsec$ radius, while background spectra used an annulus with internal radius r$_{in}$ = 6$\arcsec$ and external radius r$_{out}$ = 15$\arcsec$. Background annuli experienced no contamination from nearby sources. We applied point-source aperture correction as a part of the source spectral extraction process. We fit the spectra using Cstat, given how the low count statistic of our sources does not allow the minimum 15 cts/bin required to use $\chi^2$ statistics (for all except MCG $+$08-33-046\footnote{Even in the case of enough counts per bin, \cite{Lanzuisi2013} showed that cstat yields more constraining results ($<$ 30\% for most sources) when compared to a $\chi^2$ analysis.}). To bin the spectra, we followed \cite{Lanzuisi2013} (Appendix A), which study the optimal binning to use C-statistic (cstat) as a fitting statistic. They find that fitting is stable regardless of binning at either 1 or 5 cts/bin, but 5 cts/bin results in the optimal error determination. Therefore, we adopt this value except in two cases: ESO 090$-$IG 014 and IRAS 11058$-$1131. Due to the low net counts \textbf{($<$90 cts)} of these two sources, 5 cts/bin resulted in the dilution of visible features, such as the iron line, which affected the fit negatively. For these two sources, we opted for a 3 cts/bin; the value that is closest to 5 without erasing the iron line.


\begin{deluxetable*}{ccccccccc}
\tablecaption{Summary of Sources in Our Sample} 
\label{tab:sources}

\tablehead{\colhead{\swift-BAT ID} &  \colhead{Source Name} & \colhead{R.A.} & \colhead{Decl.} & \colhead{$z$} & \colhead{Exp. Time} & \colhead{Count Rate} & \colhead{Obs Date} & \colhead{\textbf{Type}}\\ \colhead{} & \colhead{} & \colhead{(J2000)} & \colhead{ (J2000)} & \colhead{} & \colhead{(ks)} & \colhead{(1--7\,keV)}& \colhead{}}

\startdata
J0205.2$-$0232 & 2MASX J02051994$-$0233055$^*$	& 02:05:19.9 & $-$02:33:05.9 &	0.0283 & 9.9 & 4.40e-2 & 2018 June 11 & G$^a$\\
J0402.6$-$2107 & ESO 549$-$50$^{\dagger}$ & 04:02:46.1 & $-$21:07:08.6 & 0.0252 & 9.9 & 1.55e-2 & 2019 Nov 23 & Sy2$^b$ \\
J0407.8$-$6116 & 2MASX J04075215$-$6116126$^*$ & 04:07:52.1 &	$-$61:16:12.8 & 0.0214 & 10.4 & 1.58e-2 & 2018 May 05 & G$^a$ \\
J0844.8$+$3055 & 2MASX J08445829$+$3056386$^*$ & 08:44:58.3 & 
+30:56:38.3 &	0.0643 &	9.9 & 1.15e-1 & 2018 Jan 03 & AGN$^b$ \\
J0901.8$-$6418 & ESO 090$-$IG 014$^*$	& 09:01:37.2 &$-$64:16:28.1 &	0.0220 & 9.9 & 8.29e-3 & 2018 June 01 & IG$^c$ \\
J1108.4$-$1148 & IRAS 11058$-$1131$^{\dagger}$ & 11:08:20.3 & $-$11:48:12.1 & 0.0548 & 9.7 & 3.83e-3 & 2020 Mar 2 & Sy2$^b$ \\
J1111.0$+$0054 & 2MASX J11110059$-$0053347$^{\dagger}$ & 11:11:00.6 & $-$00:53:34.9 & 0.0908 & 9.9 & 4.61e-2 & 2020 May 12 & Sy2$^b$ \\
J1258.4$+$7624 & IRAS 12571$+$7643$^{\dagger}$ & 12:58:36.0 & +76:26:41.3 & 0.0634 & 9.9 & 1.95e-2 & 2020 Jan 22 & G$^a$ \\
J1828.8$+$5021 & MCG $+$08-33-046$^{\dagger}$ & 18:28:48.1 & +50:22:20.9 & 0.0169 & 9.9 & 1.90e-1 & 2020 Jan 24 & Sy2$^b$ \\
\enddata
Notes: \\
$*$: From the 100-month BAT Catalog. \\
$\dagger$: From the 150-month BAT Catalog. \\
a: Galaxy. \cite{HyperLEDA2003} \\
b: Seyfert 2 Galaxy. \cite{VeronCetty2006} \\
c: Interacting Galaxy. \cite{Arp1987} \\

\end{deluxetable*}

\section{Spectral Analysis Results} \label{sec:all_models}
Spectral fitting was conducted with \texttt{XSPEC} v. 12.10.1f \citep{Arnaud1996}. The Heasoft tool \texttt{nh} \citep{Kalberla05} was used to calculate Galactic absorption in the direction of each source. The flux and intrinsic luminosity for each source were calculated using \texttt{clumin}\footnote{\url{https://heasarc.gsfc.nasa.gov/xanadu/xspec/manual/node281.html}} in \texttt{xspec}. The tables listing the results of the 1$-$7\,keV spectral fitting of the nine \swift-BAT galaxies are reported in the Appendix.  In this section, we introduce the models that are used to analyze the source spectra in $\S$\ref{sec:models} and the fitting results in $\S$\ref{sec:sources}.

\subsection{Models Implemented} \label{sec:models}
\subsubsection{\texttt{MYTorus}}
\indent 
The \myt \citep{Murphy2009} model assumes a uniform torus of absorbing material with circular cross section and a fixed opening angle of 60$\degr$. The model is composed of three components: the line-of-sight continuum, the Compton-scattered component (i.e. reflection), and the fluorescent lines. The line-of-sight continuum, also described as the zeroth-order continuum, is the intrinsic X-ray emission from the AGN as observed after absorption from the surrounding torus. Next, the Compton-scattered component models the photons that scatter into the observer line of sight after interacting with the dust and gas surrounding the SMBH. In the case the true covering factor of the source differs from the default \myt value, $f_c = cos(\theta) =$ 0.5, or a not negligible time delay exists between the intrinsic emission and the scattered component,
the two components require different normalizations. The normalization for the scattered component is denoted as $A_S$ \citep{Yaqoob2012}. The final component models the most significant fluorescent lines, i.e., the Fe K$\alpha$ and Fe K$\beta$ lines, at 6.4 and 7.06\,keV, respectively. This component also has its own normalization, $A_L$. $A_S$ and $A_L$ were fixed to 1 due to the low quality of our spectra. In our analysis, we also searched for the presence of a thermal component below 1\,keV, which is observed in the X-ray spectra of AGN. However, due to the low count statistic of our spectra at soft X-ray energies, none of our fits was significantly improved when adding a thermal component (\texttt{mekal}). \\ 
\indent In \texttt{XSPEC} notation, our model is defined as:

\begin{eqnarray}
 \label{eq:myt}
    ModelA = constant_1  * phabs * \nonumber (MYTZ * \\ zpowerlw + \nonumber A_S * MYTS + \nonumber A_L * MYTL \nonumber \\ + f_s * zpowerlw),
\end{eqnarray}

\noindent where $MYTZ*zpowerlw$ represents the line-of-sight continuum (or the zeroth-order continuum), $MYTS$ the scattered component, and $MYTL$ the fluorescent lines. Lastly, $f_s$ is the fraction of intrinsic emission that leaks through the torus
rather than being absorbed by the obscuring material. 
The \myt model can be utilized in two different configurations, designated `coupled' and `decoupled' \citep{Yaqoob2012}.

\subsubsection{\myt in Coupled Configuration}
\myt measures the angle between the axis of the torus and the observer line of sight, known as the torus inclination angle, which will hereafter be written as $\theta_{obs}$. This angle ranges from 0$\degr-$90$\degr$, where $\theta_{obs}$ = 90$\degr$ represents edge-on observing and $\theta_{obs}$ = 0$\degr$ represents face-on observing. When \myt is in the coupled configuration, all three components of the model are set to have the same $\theta_{obs}$, which is a free parameter, and the same column density.

\subsubsection{\myt in Decoupled Configuration}
The decoupled configuration of \myt was initially introduced in \cite{Yaqoob2012} and adds the flexibility of allowing different values for the line-of-sight column density, $N_{H,Z}$, and the average torus column density, $N_{H,S}$; a first approximation to a clumpy distribution. In this configuration, the line-of-sight continuum is fixed to an angle of $\theta_{obs,Z}$ = 90$\degr$.
The scattered and fluorescent line components have an equal $\theta_{obs,S,L}$, which can either be fixed to 90$\degr$ or 0$\degr$ to represent edge-on or face-on reflection, respectively. In the edge-on reflection scenario, the obscuring material between the AGN and the observer reprocesses the photons.
In the face-on reflection scenario, the emission reflecting off the far side of the torus passes through and is observed (which could also be a sign of a patchy distribution of material). In \myt decoupled, the scattered and fluorescent line column densities are represented by $N_{H,S}$, which can vary greatly from the line-of-sight column density in an inhomogeneous, patchy torus.

\subsubsection{BORUS02}
The second physically motivated model we used to analyze our data is \texttt{borus02} \citep{Balo_borus2018}. This model incorporates an absorbed intrinsic continuum multiplied by a line-of-sight absorbing component, \texttt{zphabs $\times$ cabs}. Additionally, \bor models the reprocessed component, including the Compton-scattered component and fluorescent lines. This model includes the torus covering factor as a free parameter which varies in the range from $f_c$ = 0.1$-$1, equivalent to a torus opening angle $\theta_{tor}$ = 0$\degr-$84$\degr$. 
\texttt{borus02} is implemented in \texttt{XSPEC} in the following way:

\begin{equation}
\begin{aligned}
 \label{eq:borus}
    ModelB = constant_1  * phabs * \\ (borus02 + zphabs * cabs * zpowerlw \\
    + f_s * zpowerlaw),
\end{aligned}
\end{equation}

\noindent where $borus02$ models the reprocessed components, which includes the scattered continuum and fluorescent line emission. In addition, \bor includes the high-energy cutoff and iron abundance as free parameters. We froze the energy cutoff at 500\,keV to remain consistent with the default setting in \myt and fixed the iron abundance to 1 due to low statistics in our data. The \textit{zphabs} and \textit{cabs} components account for the line-of-sight absorption 
within the source, including Compton scattering losses out of the line of sight. \\

\section{Results} \label{sec:sources}
\begin{figure}
    \hspace{-0.5cm}
    \includegraphics[scale=0.45]{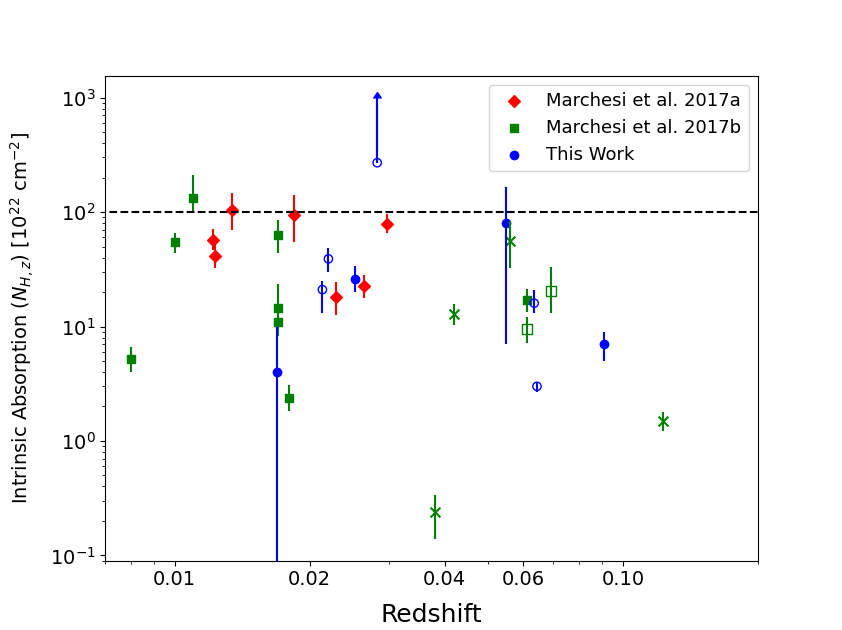}
    \caption{The line-of-sight column density as a function of redshift. The blue circles represent the \bor best fit line-of-sight column densities from the nine sources in the work, the red diamonds represent the seven sources analyzed in \cite{Marchesi17a} with a fixed inclination angle of 90$\degr$, and the green squares represent the results from \cite{Marchesi_2017b}. Unfilled shapes are reported as galaxies in SIMBAD (see Table \ref{tab:sources}). The four X symbols are Seyfert 1 galaxies. Sources above the dashed line are candidate Compton-thick AGNs (i.e., having line-of-sight column density N$_{H,Z}$ $\geq$ 10$^{24}$ cm$^{-2}$). The arrow pointed upwards represents a source with only a lower limit on the line-of-sight N$_{H,Z}$.}
    \label{fig:nh}
\end{figure}

In Figure \ref{fig:nh}, we report the \texttt{borus02} best-fit line-of-sight column density as a function of redshift for the nine objects presented in this paper, as well as for those obtained in previous works by our group. The subsequent subsections describe the fitting results for all nine sources. For every source, we removed from the BAT spectra those data points with error bars compatible with a non-detection. We note that in the tables of best-fit parameters for the final five sources, we only list the \textit{Chandra} cstat/dof as the BAT data showed poor fitting statistics, likely due to the large intrinsic dispersion. For example, the BAT data for 2MASX J11110059$-$0053347 had only 10 points yet was responsible for $\sim$90\% of the reduced statistic. \\
\indent We note that we compared our errors from the model fits with errors calculated using a Markov Chain Monte Carlo (MCMC) algorithm\footnote{\url{https://heasarc.gsfc.nasa.gov/xanadu/xspec/manual/node43.html}}. As both methods produced similar results, we are confident the errors derived from the models are valid. \\
\indent In the Appendix, we show the analysed \textit{Chandra} images of each source, along with the corresponding BAT positional uncertainty region at a 95\% confidence level. In each image, the counterpart listed in Table \ref{tab:sources} is marked within the BAT region. The region size is calculated by 

\begin{equation}
   R_{95}(\arcmin) = 12.5 \times [S/N]^{-0.68} + 0.54,
\end{equation}

where S/N is the significance of the detection \citep[see][for details]{Cusumano2010, Segreto2010}. The size of this region for each source can be found in Table \ref{tab:batregions}. For the majority of our sources, there is only one object within the BAT region, making it the clear counterpart. In cases where it is not as clear, further discussion can be found below, in the respective subsection detailing the analysis of each object. \\
\indent We note that the \textit{Chandra} and BAT data are not taken simultaneously, as BAT gives an average of the spectrum over 150 months of observation. This implies our observations are susceptible to being affected by variability. Any \textbf{non-CT} variability in the line-of-sight column density does not have an impact in our analysis, given how BAT is sensitive to emission $> 15$\,keV, an energy range unaffected by line-of-sight obscuration. However, flux variability between the two observations can occur, and therefore we have included a \textit{Chandra} cross-normalization constant, C$_{cha}$, to account for this. It is possible that extreme flux variability can occur, causing our models to mistakenly classify a source as highly obscured when it is only in a low-flux state. We note that this has not happened before and our previous works show that the \textit{Chandra}$-$BAT analysis produces reasonably accurate column density measurements (however with large uncertainties) when compared with the \textit{XMM-Newton}$-$\textit{NuSTAR} results \citep{Marchesi17a, Zhao2019B, Zhao2019A}. This is why any of our sources that are compatible with $N_{H}>10^{24}$ cm$^{-2}$ must be considered only a \textit{candidate} CT-AGN until simultaneous \textit{XMM-Newton} and \textit{NuSTAR} observations can verify the classification.


\subsection{\textbf{X-ray Spectral Fitting Results}}

\subsubsection{ESO 090$-$IG 014} 

\begin{figure*}
    \centering
    \includegraphics[scale=0.11, clip=true,trim=200 100 100 100]{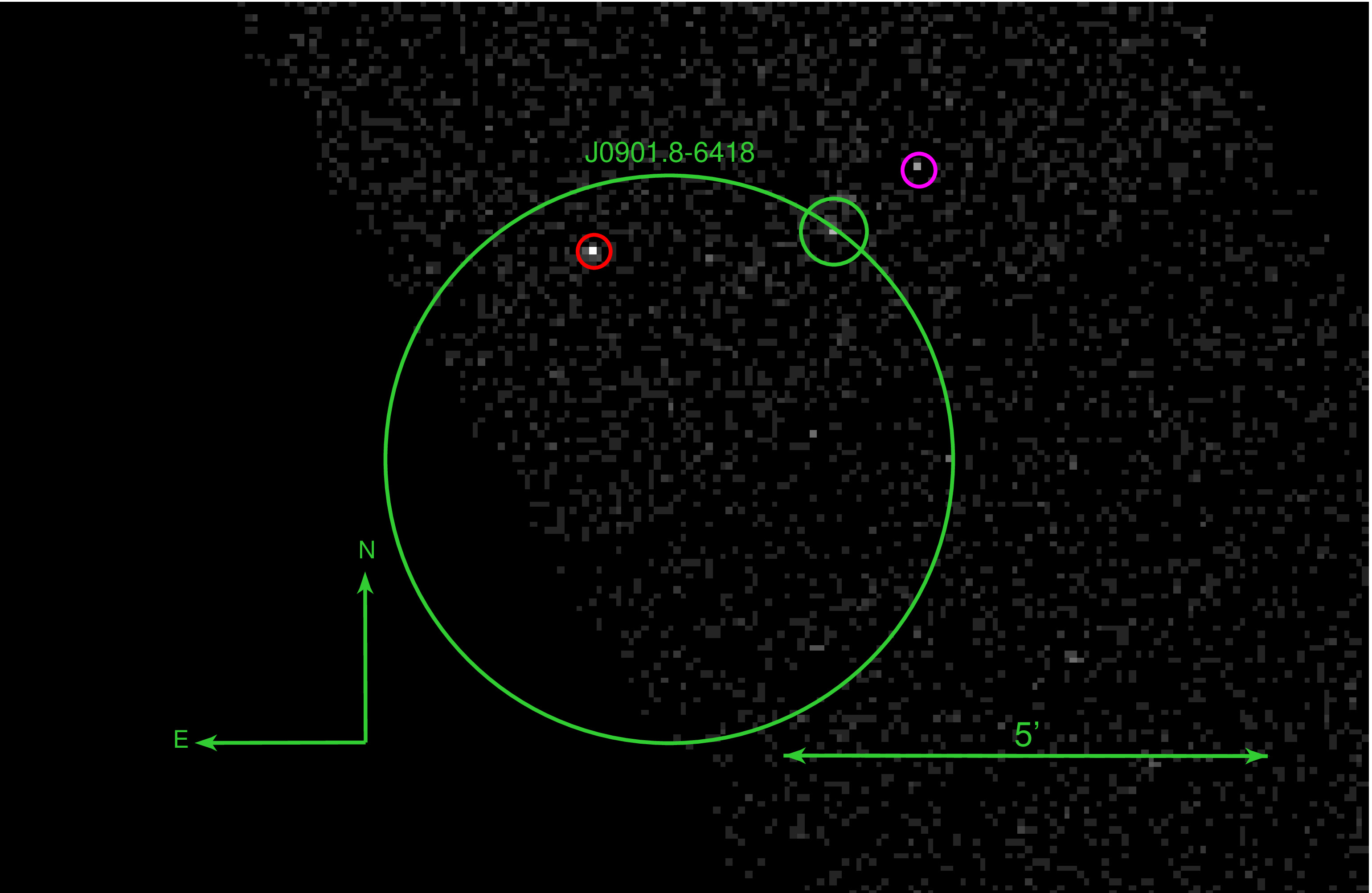}
    \caption{The \textit{Chandra} CCD for the 10\,ks of ESO 090$-$IG 014 (\textbf{marked with the smaller of the two green circles}) \textbf{in the 1$-$7\,keV range}. Two other nearby sources, 2MASS J09015969$-$6416408 (red) and WISEA J090129.46$-$641551.1 (magenta), were studied and are believed to have not contributed to the BAT emission \textbf{(see the text for further details)}. \textbf{The large green circle (2.87$\arcmin$ in radius, see Table \ref{tab:batregions}) represents the 95\% \textit{Swift}-BAT confidence region.}}
    \label{fig:eso090}
\end{figure*}

As is visible in Figure \ref{fig:eso090}, there are 2 other X-ray sources in the \textit{Chandra} field, 2MASS J09015969$-$6416408 (red) and WISEA J090129.46$-$641551.1 (magenta) that are within or near ($< 1\arcmin$) the BAT 95\% confidence region (R $\approx$ 3$\arcmin$). Neither source has any X-ray emission above 3\,keV. Furthermore, ESO 090$-$IG 014 (green) is $\sim$3 magnitudes brighter in the WISE W3 band, the band most commonly associated with AGN emission \citep{Asmus015}. For these reasons, it is highly unlikely that either of the other two sources would contribute to the \textit{Swift}-BAT data. \\
\indent The best fit results are displayed in Table \ref{tab:eso09}. All four  models show good agreement with a soft photon index $\Gamma \approx$ 2.20. ESO 090$-$IG 014 is one of the sources for which the best fit required a cross-normalization not equal to 1, C$_{cha}\approx0.3$, suggesting that the \textit{Chandra} observation was taken in a low-flux epoch. Except \myt coupled, all models suggest an obscured AGN with $N_{H,Z} \approx 4\times10^{23}$ cm$^{-2}$. The average torus column density is lower, on the order of 10$^{22}$ cm$^{-2}$, suggesting a clumpy torus. In the \bor model, the covering factor and $\theta_{obs}$ were fixed to 0.5 and 87$\degr$ (the upper limit in \texttt{borus02}), respectively, since the data quality was not high enough to properly constrain them \citep[as suggested by][]{Zhao2020}. We note these values are consistent with the poorly-constrained best-fit values. Also, $f_s$ is frozen to zero in \texttt{borus02}, given how the best-fit results yields $f_s$ $<$ 10$^{-5}$, compatible with zero.

\begin{deluxetable*}{ccccc}

\tablecaption{ESO 090$-$IG 014}
\label{tab:eso09}

\tablehead{\colhead{Model} & \colhead{MYTorus} & \colhead{MYTorus} & \colhead{MYTorus} & \colhead{borus02} \\ 
\colhead{} & \colhead{(Coupled)} & \colhead{(Decoupled Face-on)} & \colhead{(Decoupled Edge-on)} & \colhead{}}
\startdata
cstat/dof & 43/27 & 42/27 & 42/27 & 40/27 \\
$\Gamma$ & 2.19$^{+0.17}_{-0.18}$ & 2.17$^{+0.17}_{-0.17}$ & 2.21$^{+0.16}_{-0.17}$ & 2.12$^{+0.16}_{-0.18}$ \\
$N_{H,eq}$ & 1.35$^{+1.90}_{-1.03}$ & ... & ... & ... \\
norm $10^{-2}$ & 0.42$^{+0.09}_{-0.07}$ & 0.50$^{+0.14}_{-0.09}$ & 0.60$^{+0.08}_{-0.06}$ & 0.44$^{+0.02}_{-0.02}$ \\
c$_{f,Tor}$ & ... & ... & ... & 0.5* \\
cos($\theta_{obs}$) & 0.49$^{+}_{-0.01}$ & ... & ... & 0.05* \\
$N_{H, Z}$ & ... & 0.37$^{+0.12}_{-0.10}$ & 0.42$^{+0.16}_{-0.08}$ & 0.40$^{+0.10}_{-0.09}$ \\
$N_{H, S}$ & ... & 0.01$^{+0.25}_{-}$ & 0.03$^{-}_{-}$ & 0.01$^{+0.23}_{-}$ \\ 
$f_s$ 10$^{-2}$ & 0.20$^{+0.23}_{-0.17}$ & 0.24$^{+0.33}_{-}$ & 0.07$^{+0.26}_{-}$ & 0* \\
C$_{cha}$ & 0.34$^{+0.21}_{-0.14}$ & 0.30$^{+0.17}_{-0.11}$ & 0.32$^{+0.20}_{-0.17}$ & 0.34$^{+0.20}_{-0.06}$ \\
\hline
L$\rm_{2-10\,keV}$ & & & & 1.02$^{+0.24}_{-0.19}$ $\times$ 10$^{43}$\\
L$\rm_{15-55\,keV}$ & & & & 6.46$^{+0.62}_{-0.57}$ $\times$ 10$^{42}$ \\
\enddata
\vspace{0.5cm}
\textbf{Notes:} \\ 
 *: Parameter was frozen to this value during fitting. \\
 $^{-}_{-}$: Parameter is unconstrained. \\
 $\Gamma$: Power law photon index. \\
 N$_{H,eq}$: Hydrogen column density along the equator of the torus in units of 10$^{24}$ cm$^{-2}$. \\
 norm: the main power-law normalization (in units of photons cm$^2$ s$^{−1}$ keV$^{−1}$ $\times$ 10$^{−4}$), measured at 1 keV. \\
 c$_{f,Tor}$: Covering factor of the torus. \\
 cos($\theta_{obs}$): Cosine of the inclination angle. \\
 $N_{H, Z}$: Line-of-sight torus hydrogen column density, in units of 10$^{24}$ cm$^{−2}$. \\
$N_{H, S}$: Average torus hydrogen column density, in units of 10$^{24}$ cm$^{−2}$. \\
$f_s$: Fraction of scattered continuum. \\
C$_{cha}$: The cross-normalization constant between the \textit{Chandra} and \textit{Swift}-BAT data. \\
L$\rm_{2-10\,keV}$: Intrinsic luminosity in the 2$-$10\,keV band with units of erg s$^{-1}$. \\
L$\rm_{15-55\,keV}$: Intrinsic luminosity in the 15$-$55\,keV band with units of erg s$^{-1}$.
 
\end{deluxetable*}

\subsubsection{2MASX J02051994$-$0233055}

The 150-month catalog lists 2MASX J02051994$-$0233055 (green in Figure \ref{fig:field1}) as the counterpart of the BAT emission. The 105-Month \textit{Swift}-BAT catalog \citep{Oh2018}, however, includes a BAT source that overlaps with the one studied here, with WISEA J020527.94$-$023321.8 (red) listed as its counterpart. We mark the position of both possible counterparts in the \textit{Chandra} field image in Appendix \ref{AppB}, showing that there is no \textit{Chandra} emission coming from WISEA J020527.94$-$023321.8. Moreover, WISEA J020527.94$-$023321.8 is $\sim$4 magnitudes dimmer in the W3 band. The counterpart of the BAT emission is thus 2MASX J02051994$-$0233055.

\indent While the majority of the models for 2MASX J02051994$-$0233055 are in agreement, the \myt decoupled edge-on configuration yielded significantly different results in the line-of-sight column density (8.6$^{+4.0}_{-2.5}$ $\times$ 10$^{23}$ cm$^{-2}$). The other three models suggest a heavily CT-AGN with line-of-sight column density N$_{H,Z} >>$ 10$^{24}$ cm$^{-2}$.
According to the \bor model results, displayed in Figure \ref{fig:spec_cont1}, the spectrum is reflection dominated. \\
\indent We tested the reliability of the reflection-dominated scenario using the pexmon model as follows:\\

\begin{equation}
\begin{aligned}
    ModelC = constant * phabs * (zphabs * zpowerlw + \\ pexmon + f_s * zpowerlw).
\end{aligned}
\end{equation}
\texttt{pexmon} is a neutral Compton reflection model with self-consistent Fe and Ni lines \citep{George1991, Nandra2007}. It includes a scaling factor, R, which allows to consider (R = 1) or exclude (R $<$ 0) the intrinsic emission component. We opt for the latter, as the inclusion of \textit{zphabs*zpowerlaw} allows to estimate the line of sight column density. The best fit line-of-sight column density for the \texttt{pexmon} model is N$_{H,Z}$ = 5.54$^{+5.76}_{-3.44}$ $\times$ 10$^{24}$ cm$^{-2}$. \\
\indent Finally, we also attempt to model the source using the \myt decoupled model in a combination of both face-on and edge-on configurations. This model is consistent with the \texttt{MYTorus} coupled, \myt decoupled face-on, and \bor results, with best-fit estimations all suggesting line-of-sight and average column densities on the order of 10$^{25}$ cm$^{-2}$. Based on the results derived from all models, we believe 2MASX J02051994$-$0233055 to be a reliable CT-AGN candidate.

\subsubsection{2MASX J04075215$-$6116126}
ESO 118$-$IG 004 (red in Figure \ref{fig:field1}) was originally listed as the most reliable counterpart for 4PBC J0407.8$-$6116 as it is bright in the W3 band ($\sim$7.5). However, the analysis of the \textit{Chandra} data allowed us to challenge this claim, and determine that 2MASX J04075215$-$6116126 \textbf{(green)} is the true source of the BAT emission. As seen in Appendix \ref{AppB}, 2MASX J04075215$-$6116126 is at the center of the BAT 95\% confidence region, while ESO 118$-$IG 004 is located just outside of it. Also, 2MASX J04075215$-$6116126 has a count rate (in the 2$-$10\,keV band) more than 15 times greater than that of ESO 118$-$IG 004. Furthermore, ESO 118$-$IG 004 shows a flux decay toward higher energies, while 2MASX J04075215$-$6116126 shows an increase consistent with a Compton-thin source. However, it is not impossible that ESO 118$-$IG 004 is a highly CT source that contributes in a non-negligible way to the BAT flux. For this reason, we have been awarded NuSTAR + XMM simultaneous observations to study these sources and verify their column densities. \\
\indent We note that the source WISEA J040732.53$-$611918.3 (magenta) is also present in the \textit{Chandra} image, located just outside of the BAT confidence region. However, it is $\sim$5 times dimmer in X-rays and $\sim$3 magnitudes dimmer in the W3 band. Therefore, we do not believe it contributes to the BAT flux. \\
\indent The four models used agree on most parameters. All estimate a photon index $\Gamma \approx$ 1.60 with only 7\% errors and a line-of-sight column density around 2.0$\times$10$^{23}$ cm$^{-2}$. The reflection component parameters are less tightly constrained due the fact that, in this particular source, reflection is subdominant. The average torus column density is similar to the line-of-sight value, $\approx$ 2.0$\times$10$^{23}$ cm$^{-2}$, however with uncertainties greater than half an order of magnitude. The \bor model yields a value of 0.90 for both the covering factor and the cosine of the inclination angle although largely unconstrained.

\subsubsection{2MASX J08445829$+$3056386}

2MASX J08445829$+$3056386 was also detected in the 105-Month BAT catalog under the counterpart name FBQS J084458.3+305638. \\
\indent In this work, all four models are in good agreement with most parameters, including an average AGN photon index of $\Gamma \approx$ 1.75. The two \myt decoupled models and \bor suggest this is a Compton-thin AGN with an N$_{H,Z}$ of the order of 10$^{22}$ cm$^{-2}$. This result is consistent with the fact that the source has a high count rate, 0.115 ct s$^{-1}$, which would be unusual for a \textit{Swift}-BAT-detected CT-AGN at $z \sim 0.07$. The \bor model cannot constrain $\theta_{obs}$, so it was fixed to cos($\theta_{obs}$) = 0.05. In addition, the scattering constant $f_s$ is also fixed to zero due to its best-fit result being compatible with zero, i.e., $f_s <$ 10$^{-5}$.

\subsubsection{2MASX J11110059$-$0053347}
\indent All four models displayed in Table \ref{tab:2m111} show consistent N$_{H,Z}$ values $\sim$ 7$\times$10$^{22}$ cm$^{-2}$, signaling a Compton-thin AGN. The photon index was well constrained around $\Gamma$=1.65 with $\sim$5\% errors. As evidenced by the low obscuration measured for this source, we conclude the intrinsic emission dominates over the reflection component. Therefore, we were unable to provide any constraint on the torus parameters derived from the reflection component such as the average torus column density, the inclination angle, and the covering factor.

\subsubsection{ESO 549$-$50}
All models yielded a well-constrained average photon index of $\Gamma \approx$ 1.85 with only 10\% uncertainties. Moreover, the four best-fit results are in agreement with an N$_{H,Z}$ $\approx$ 3$\times$10$^{23}$ cm$^{-2}$, indicating this source is significantly obscured. While the covering factor appears to be well constrained around 0.35, neither the inclination angle nor the average torus column density are. The average torus column density is only loosely constrained (N$_{H,tor} >$ 10$^{23}$).

\subsubsection{IRAS 11058$-$1131}
The \textit{Chandra} image in Appendix \ref{AppB} shows IRAS 11058$-$1131 surrounded by several hot pixels, as well as the source 2MASS J11083339$-$1151500 (red in Figure \ref{fig:field2}) near the bottom of the BAT region. Upon further inspection, it was revealed the latter does not have emission above 3\,keV and is $\sim$ 4 magnitudes fainter in the WISE W3 band. Thus, is unlikely to be the true BAT counterpart. \\
\indent As can be seen in Figure \ref{fig:spec_cont3}, this source has a significant drop-off in flux in the \textit{Chandra} data compared to the BAT data. For this reason, we let the \textit{Chandra} cross-normalization constant, $C_{cha}$, free to vary and fixed the BAT constant to 1. Most of the models have a $C_{cha}$ value $< 1$, although with large uncertainties. IRAS 11058$-$1131 is best-fit with a high N$_{H,Z}$ value as exemplified by the large decrease in the soft-energy band. These values range from Compton-thin, 8$\times$10$^{23}$ cm$^{-2}$, to CT, 2.6$\times$10$^{24}$ cm$^{-2}$. Since both of these results are statistically equivalent, we list IRAS 11058$-$1131 as a CT-AGN candidate. The average torus column density shows better agreement among the three applicable models with values in the CT regime. The photon index is rather hard, averaging $\Gamma\sim$1.43. 

\subsubsection{MCG $+$08-33-046}
The \textit{Chandra} observation of MCG $+$08-33-046 is significantly affected by pile-up \citep[over 20\% according to the \textit{Chandra} pileup tool,][]{pileup2001}. Considering this tool is only applicable to a power law, we were unable to utilize the models discussed in Section \ref{sec:models}. With the pileup tool implemented, the best-fit has a photon index $\Gamma\sim$2 and an N$_{H,Z}$ $\approx$ 2$\times$10$^{22}$ cm$^{-2}$. This low line-of-sight column density agrees well with the high flux levels of this source in the soft X-rays.

\subsubsection{IRAS 12571$+$7643}
We originally assumed PGC 044558 (red Figure \ref{fig:field3}) as the most likely counterpart of the BAT source, due to its Sy2 nature. However, there is no emission from this source in the \textit{Chandra} exposure. As there is significant emission from IRAS 12571+7643 (green), which is also within the BAT 95\% confidence region, we believe this is the true source of the BAT emission. \\
\indent This is another Compton-thin candidate, with N$_{H,Z}$ $\approx$ 1.7$\times$10$^{23}$ cm$^{-2}$. Most fits yielded an average column density on the order of 10$^{22}$ cm$^{-2}$, significantly less than the line-of-sight N$_H$. This is only the third source in this sample to have N$_{H,Z}$ $>$ N$_{H,S}$. The photon index was consistently around $\Gamma\sim$1.7 with $<$ 8\% uncertainties.

\subsection{Mid-IR Comparison}  
\begin{figure}
    \hspace{-1.4cm}
    \includegraphics[scale=0.6]{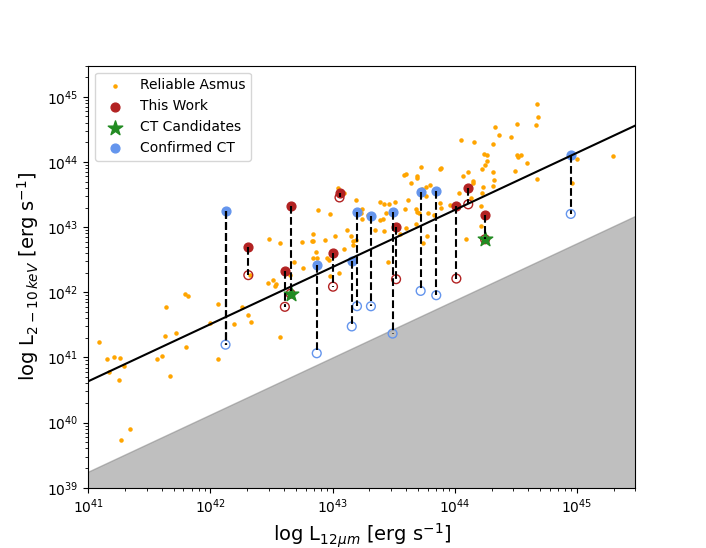}
    \caption{\textbf{The 2$-$10\,keV vs 12$\mu$m luminosities of the nine sources in this work. The filled circles represent the intrinsic 2$-$10\,keV luminosities, while the open circles represent the observed luminosities. The green stars are the two candidate CT-AGN presented in this work, 2MASX J02051994$-$0233055 and IRAS 11058$-$1131. The black line is the relation derived from the X-ray observations of a reliable sample of 152 AGN presented in \cite{Asmus015}. The blue circles are recently confirmed CT-AGN \citep{Marchesi2019, Zhao2019B, Zhao2019A, TorresAba2021, Traina2021}. The grey area represents a 25$\times$ (or more) decrease in X-ray flux, a diagnostic used to identify CT-AGN \citep{Annuar2020}. The 12$\mu$m data were obtained by WISE.}}
    \label{fig:lum_plot}
\end{figure}

\textbf{The Mid-infrared (MIR) is another useful waveband to select CT-AGN candidates. As the ultraviolet (UV) emission from the accretion disk gets absorbed by the dusty torus, it becomes heated to temperatures of several hundred Kelvins. As a result, the dust radiates thermally, with its emission peaking in the MIR \citep[$\sim$3 - 30 $\mu$m,][]{Asmus2020}. As this emission is much less susceptible to absorption, many works have used the MIR to identify heavily obscured AGN \citep{Yan2019, Asmus2020, Kilerci2020, Guo2021}. Moreover, \cite{Asmus015} used a sample of 152 AGN with reliable soft X-ray data to model a trend between the intrinsic 2$-$10\,keV luminosity and the 12$\mu$m luminosity (see Figure \ref{fig:lum_plot}). In addition, we have plotted the intrinsic (closed circles) and observed (open circles) X-ray luminosities of the sources in this work, and those of recently confirmed CT-AGN \citep{Marchesi2019, Zhao2019B, Zhao2019A, TorresAba2021, Traina2021}. The two green stars represent the candidate CT-AGN presented in this work, 2MASX J02051994$-$0233055 and IRAS 11058$-$1131. It can be seen, 2MASX J02051994$-$0233055 shows a significant decrease from intrinsic to observed luminosity and IRAS 11058$-$1131 has an observed luminosity near the CT region. Both are strong indicators of a CT-AGN.}

\section{Study of Extended Emission} \label{sec:ext_emi}
\begin{figure}
    \hspace{-0.65cm}
    \includegraphics[scale=0.6]{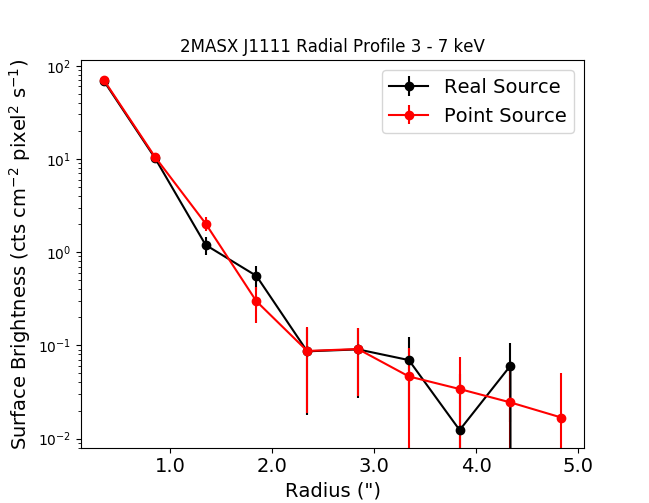}
    \caption{The 3$-7$\,keV radial profile of 2MASX J11110059$-$0053347 compared to that of a simulated point source. The two curves are compatible, and therefore there is no significant evidence of extended emission. A similar trend exists for the other eight sources in this paper.}
    \label{fig:rad_prof}
\end{figure}

The radiation from the accretion disk of an AGN is collimated by the torus, given how it is symmetric around the accretion flow axis. The radiation escaping from the AGN excites the gas of the interstellar medium (ISM) by photoionization, which appears in the form of cones extending from the nucleus. These cones have been observed in local Seyfert galaxies in NIR and optical \citep[e.g.][]{Durre2018}, as well as X-rays \citep{Fabbiano2017, Fabbiano2018, Jones2020, Ma2020}. In X-rays, the mentioned works have used \textit{Chandra's} unmatched resolution to observe kiloparsec-scale diffuse emission in both the hard continuum (3$-$7 keV) and in the Fe-K$\alpha$ line. Obscured, and particularly CT-AGN, are ideal to observe this extended X-ray emission given how the torus dims the much brighter nuclear emission. \\
\indent We take advantage of our high-resolution \textit{Chandra} images and attempt to detect the cone emission in the highly obscured sources presented in this work. In order to do so, we extract radial profiles from all of our sources and compare them to the expected radial profile of a point source of the same flux and in the same position. We follow the CIAO Point-Source Functions (PSF) simulation thread\footnote{\url{https://cxc.harvard.edu/ciao/threads/psf.html}} \footnote{\url{https://cxc.harvard.edu/ciao/threads/marx\_sim/}} and generate the \textit{Chandra} PSFs using ChaRT \footnote{\url{https://cxc.harvard.edu/ciao/PSFs/chart2/}} and MARX 5.5.1\footnote{\url{https://cxc.harvard.edu/ciao/threads/marx/}}.
Figure \ref{fig:rad_prof} shows a comparison between a simulated PSF and the observed emission, for the case of 2MASX J11110059$-$0053347. The two curves are compatible with each other, and there is no significant excess over the the simulated data counts. All sources in our sample show similar curves, and thus no sign of extended X-ray emission. 2MASX J11110059$-$0053347 has a count rate twice as high as the 10\,ks exposure of MKN 573, a CT-AGN source analyzed in \citep{Jones2021}, which presents significant extended emission. Therefore, the ionization cone in the sources of our sample is either much fainter, or not present.

\section{Discussion and Conclusions} \label{sec:disc}
In this paper, we presented the joint \textit{Chandra}--\textit{Swift}-BAT spectral fitting analysis in the 1$-$150\,keV energy range for nine nearby ($z <$ 0.1) AGN selected from the 150-month \textit{Swift}-BAT all-sky survey catalog. This represents the second step of our three step plan to discover new Compton-thick AGN in the local Universe. Our first step was selecting these sources following previous successful selection criteria \cite[][and see Sect. \ref{sec:data_reduc}]{Marchesi17a} and acquiring \textit{Chandra} snapshot observations for each of them. The third and final step will involve obtaining and analyzing XMM-\textit{Newton} and \textit{NuSTAR} observations of the best CT-AGN candidates found in this work. We identified these candidates by fitting the \textit{Swift}-BAT and \textit{Chandra} spectra  with several models in order to constrain spectral parameters such as the intrinsic absorption, N$_{H,Z}$, and photon index, $\Gamma$, to uncover highly obscured AGN. The \bor best-fit parameters for each source are listed in Table \ref{tab:all}.

\begin{center}

\begin{longrotatetable}
\begin{deluxetable*}{cccccccccc} 

\tablecaption{Results for the Entire Sample}
\label{tab:all}

\tablehead{\colhead{Source} & \colhead{ESO 090} & \colhead{2MASX J0205} & \colhead{2MASX J0407} & \colhead{2MASX J0844} & \colhead{2MASX J1111} & \colhead{ESO 549} & \colhead{IRAS 110} & \colhead{MCG $+$08} & \colhead{IRAS 125} }

\startdata
cstat/dof & 40/27 & 57/72 & 57/31 & 201/170 & 98/74 & 21/23 & 11/6 & 86/77 & 41/30 \\
$\Gamma$ & 2.12$^{+0.16}_{-0.18}$ & 1.65$^{+0.09}_{-0.22}$ & 1.56$^{+0.09}_{-0.14}$ & 1.74$^{+0.07}_{-0.06}$ & 1.77$^{+0.02}_{-0.19}$ & 1.91$^{+0.02}_{-0.02}$ & 1.45$^{+0.35}_{-}$ & 2.05 & 1.61$^{+0.18}_{-0.07}$ \\
$N_{H,eq}$ & ... & ... & ... & ... & ... & ... & .. & 0.02 & ... \\
norm $10^{-2}$ & 0.44$^{+0.02}_{-0.02}$ & 0.27$^{+1.23}_{-0.16}$ & 0.04$^{+0.02}_{-0.01}$ &  0.09$^{+0.01}_{-0.01}$ & 0.06$^{+0.02}_{-0.01}$ & 0.10$^{+0.10}_{-0.02}$ & 0.04$^{+0.16}_{-0.02}$ & 0.32 & 0.05$^{+0.02}_{-0.01}$ \\
c$_{f,Tor}$ & 0.5* & 0.87$^{+0.11}_{-0.46}$  & 0.90$^{-}_{-0.85}$ & 1.00$^{-0.65}_{-}$ & 0.89$^{-}_{-}$ & 0.35$^{+0.05}_{-0.05}$ & 0.60$^{-}_{-}$ & ... & 0.15$^{-}_{-}$ \\
cos($\theta_{obs}$) & 0.05* & 0.78$^{+0.21}_{-0.51}$ & 0.90$^{-}_{-}$ &  0.05* & 0.89$^{-}_{-}$ & 0.95$^{-}_{-0.77}$ & 0.05* & ... & 0.90$^{-}_{-}$ \\
$N_{H, Z}$ & 0.40$^{+0.10}_{-0.09}$$\dagger$ & 10.00*$\dagger$ &  0.21$^{+0.04}_{-0.08}$ &  0.030 $^{+0.003}_{-0.003}$$\dagger$  & 0.07$^{+0.02}_{-0.02}$ & 0.26$^{+0.08}_{-0.06}$ & 0.81$^{+0.84}_{-0.74}$ & ... & 0.16$^{+0.05}_{-0.03}$ \\
$N_{H, S}$ & 0.01$^{+0.23}_{-}$ & 10.00$^{-}_{-6.45}$ & 0.19$^{+2.32}_{-0.15}$ & 0.24$^{+0.22}_{-0.15}$ & 31.62$^{-}_{-}$ & 14.13$^{-}_{-}$ & 7.41$^{-}_{-}$ & ... & 0.01$^{-}_{-}$ \\
$f_s$ 10$^{-2}$ & 0* & 4.08$^{+11.90}_{-2.63}$ & 2.03* & 0* & 0.03$^{+0.04}_{-0.03}$ & 0* & 0.03$^{-}_{-}$ & ... & 0.01$^{+0.01}_{-}$ \\
C$_{cha}$ & 0.34$^{+0.20}_{-0.06}$ & 1* & 1* & 1* & 1* & 1* & 0.70$^{+5.30}_{-0.69}$ & 1* & 1* \\
\hline
L$\rm_{2-10\,keV}$ & 1.02 $\times$ 10$^{43}$ & 1.10 $\times$ 10$^{42}$ & 1.51 $\times$ 10$^{42}$ & 3.31 $\times$ 10$^{43}$ & 3.98 $\times$ 10$^{43}$ & 3.98 $\times$ 10$^{42}$ & 1.55 $\times$ 10$^{43}$ & 4.90 $\times$ 10$^{42}$ & 2.09 $\times$ 10$^{43}$ \\
L$\rm_{15-55\,keV}$ & 6.46 $\times$ 10$^{42}$ & 3.98 $\times$ 10$^{42}$ &  3.16 $\times$ 10$^{42}$ & 3.89 $\times$ 10$^{43}$ & 5.01 $\times$ 10$^{43}$ & 3.72 $\times$ 10$^{42}$ & 3.16 $\times$ 10$^{43}$ & 3.55 $\times$ 10$^{42}$ & 3.63 $\times$ 10$^{43}$ \\
\enddata
\vspace{5mm}
\textbf{Notes:} Same as Table \ref{tab:eso09}. \\ 
$\dagger$: At least one of the models has a significantly different best-fit $N_{H, Z}$ value.

\end{deluxetable*}
  \end{longrotatetable}

\end{center}

\subsection{Model Comparison}

\indent The two configurations of \texttt{MYTorus} decoupled, face-on and edge-on, and \bor were capable of satisfactorily fitting all sources, while \texttt{MYTorus} coupled yielded agreeing values in most cases. As previously discussed, the poor data quality in this sample required us to freeze multiple parameters (typically covering factor and/or inclination angle) in \texttt{borus02}, and to use a simplified version of \texttt{MYTorus} decoupled (adopting either an edge-on or face-on scenario, instead of a combination of both). While these simplifications, and the low count statistics, do not allow us to use the model complexity to estimate average torus properties, they accomplish the main goal of this paper: providing an estimate of the line-of-sight column density. This allows us to classify them as either candidate CT-AGN, or as likely C-thin sources. Here we discovered two new CT-AGN candidates, 2MASX J02051994$-$0233055 and IRAS 11058$-$1131.

For two sources we implemented additional models, to ascertain their nature. For 2MASX J02051994$-$0233055 we used a phenomenological model (\texttt{pexmon}) to confirm the dominance of the reflection component. For MCG $+$08$-$33$-$046 we used only a simple absorbed power law, given how \texttt{XSPEC} provides a tool to treat pile-up that is applicable only to a power law \citep[pileup\_map][]{pileup2001}. \\
\indent Given all the mentioned limitations, and the need to freeze the parameters that constrain the main torus properties, we have opted not to implement more complex models \citep[i.e. \texttt{UXClumpy}, which models a clumpy torus scenario,][]{Buchner2019}. We leave this interesting possibility for a follow-up project, using joint \textit{XMM-Newton} and \textit{NuSTAR} observations, on the two newly discovered CT-AGN candidates (Silver et al. in prep.).

\subsection{Efficiency of Selection Criteria}

\indent We selected nine high-latitude ($|b| >$ 10$\degr$) sources from the BAT 150 month catalog that lacked a ROSAT counterpart (0.1 $-$ 2.4\,keV) and are classified as galaxies or Seyfert 2 galaxies.
As discussed in Section \ref{sec:sources}, all nine sources exhibit some level of obscuration, with a line-of-sight Hydrogen column density $\geq$ 10$^{22}$ cm$^{-2}$. However, 
three of the sources analysed in this work have logN$_{H,Z}$ $<$ 23, while those analysed by \cite{Marchesi17a} were all above this threshold. According to the selection criteria used, the lack of a \textit{ROSAT} counterpart should imply an obscuration of at least logN$_{H,Z}$ $\geq$ 23 \citep{Koss2016}. We believe this increase in sources with lower levels of obscuration could be caused by our sampling of fluxes fainter than before.
In particular, the sources selected in this work are selected from the 150-month BAT catalogue (Segreto et al. in prep.), and were not detected in the previous 100-month version \citep{Cusumano_2014}. This makes them intrinsically fainter than those selected in \citet{Marchesi17a}. Furthermore, we performed simulations in WebSpec\footnote{\url{https://heasarc.gsfc.nasa.gov/webspec/webspec.html}} testing at which column densities our high $z$ ($z >$ 0.05) sources no longer became detectable by ROSAT. This occurred at column densities as low as 5 $\times$ 10$^{22}$ cm$^{-2}$, well below the 1 $\times$ 10$^{23}$ cm$^{-2}$ predicted by \cite{Ajello_2008} and \cite{Koss2016}. Therefore, the lack of ROSAT counterpart is not as predictive of heavily obscured AGN as initially assumed.


In any case, our results, together with those presented in \cite{Marchesi17a} and \cite{Marchesi_2017b}, show that within uncertainties, 29/30 sources are obscured AGN and 5 (i.e., 17 $\pm$ 7\%) of the sources selected through our previously mentioned criteria are classified as CT-AGN candidates based on the \textit{Chandra}-BAT analysis. However, the necessity of targeting local sources (with these criteria) becomes clear when comparing the best-fit results of sources with $z<$ 0.04 and $z\geq$ 0.04. At $z<$ 0.04, we see a success rate to discover CT-AGN of 4/20 (20 $\pm$ 10\%) and an average $N_{H,Z}$ = 8.95$\times$10$^{23}$ cm$^{-2}$. In contrast, at $z\geq$ 0.04 we have a success rate of 1/10 (10 $\pm$10\%) and an average $N_{H,Z}$ = 2.24$\times$10$^{23}$ cm$^{-2}$, approximately a quarter of that of the $z<$ 0.04 sources. Moreover, only 4 out of the 20 sources (20 $\pm$ 10\%) at $z<$ 0.04 have a best-fit $N_{H,Z} <$ 10$^{23}$ cm$^{-2}$, while 4 out of 10 (40 $\pm$ 20\%) do for $z\geq$ 0.04. Note that in a blind survey \citep[see, e.g.][]{Burlon2011} only about 5\% of AGN are found to be CT, suggesting these criteria remain a powerful tool to find heavily obscured, and especially CT, AGN. \\
\indent Besides redshift, an important selection criterion is the source optical classification. All 5 potential CT-AGN candidates are either Seyfert 2s or are galaxies without a reliable optical classification (i.e. galaxy, galaxy in pair, AGN; Segreto et al. in prep.). Sources that cannot be easily classified based on their optical spectra are more likely to be Type 2 AGN, given how the obscuration can hinder the detection of features needed for an accurate classification\footnote{We note that galaxies not optically classified as AGN with detected BAT emission are likely to be AGN, since their luminosities in the $>15$ keV band are $>10^{42}$\,erg s$^{-1}$.}. Moreover, none of the four Seyfert 1s in \cite{Marchesi_2017b} are Compton thick and two are the least obscured sources in the full sample of 30. 

\subsection{Future Work}
\indent This work is part of an ongoing effort to identify and characterize all \ct in the local ($z < 0.1$) universe (The Clemson Compton-Thick AGN project\footnote{\url{https://science.clemson.edu/ctagn/}}). In order to do so, we plan on:\\

\begin{itemize}
    \item Increasing the count statistics used on the most promising candidates. Two potential CT-AGN, 2MASX J02051994$-$0233055 and ESO 118$-$IG 004, have been accepted for joint XMM-\textit{Newton} and \textit{NuSTAR} observations for 30\,ks and 80\,ks, respectively (Cycle 6, PI: M. Ajello). The increased exposure time and sensitivity of the instruments will allow us to better characterize these CT-AGN candidates. 
    \item Implementing patchy torus models like \texttt{UXClumpy} utilizing the improved count statistics from XMM-\textit{Newton} and \textit{NuSTAR}.
    \item Increasing the sample size of potential CT-AGNs. With the recent release of the 150-month \textit{Swift}-BAT all-sky survey catalog, there are additional sources meeting our criteria that have never been observed by \textit{Chandra}, XMM-\textit{Newton}, or \textit{NuSTAR}. We plan to target these sources with future observations.
\end{itemize}



\acknowledgements
RS, NTA, AP, and MA acknowledge NASA funding under contracts 80NSSC20K0045, 80NSSC19K0531, and 80NSSC21K0016 and SAO funding under contracts GO0-21083X and G08-19083X. SM acknowledges funding from the the INAF ``Progetti di Ricerca di Rilevante Interesse Nazionale'' (PRIN), Bando 2019 (project: ``Piercing through the clouds: a multiwavelength study of obscured accretion in nearby supermassive black holes'').

\appendix 

\section{Best Fit Parameters}

\begin{figure*}
    \centering
    \includegraphics[scale=0.8]{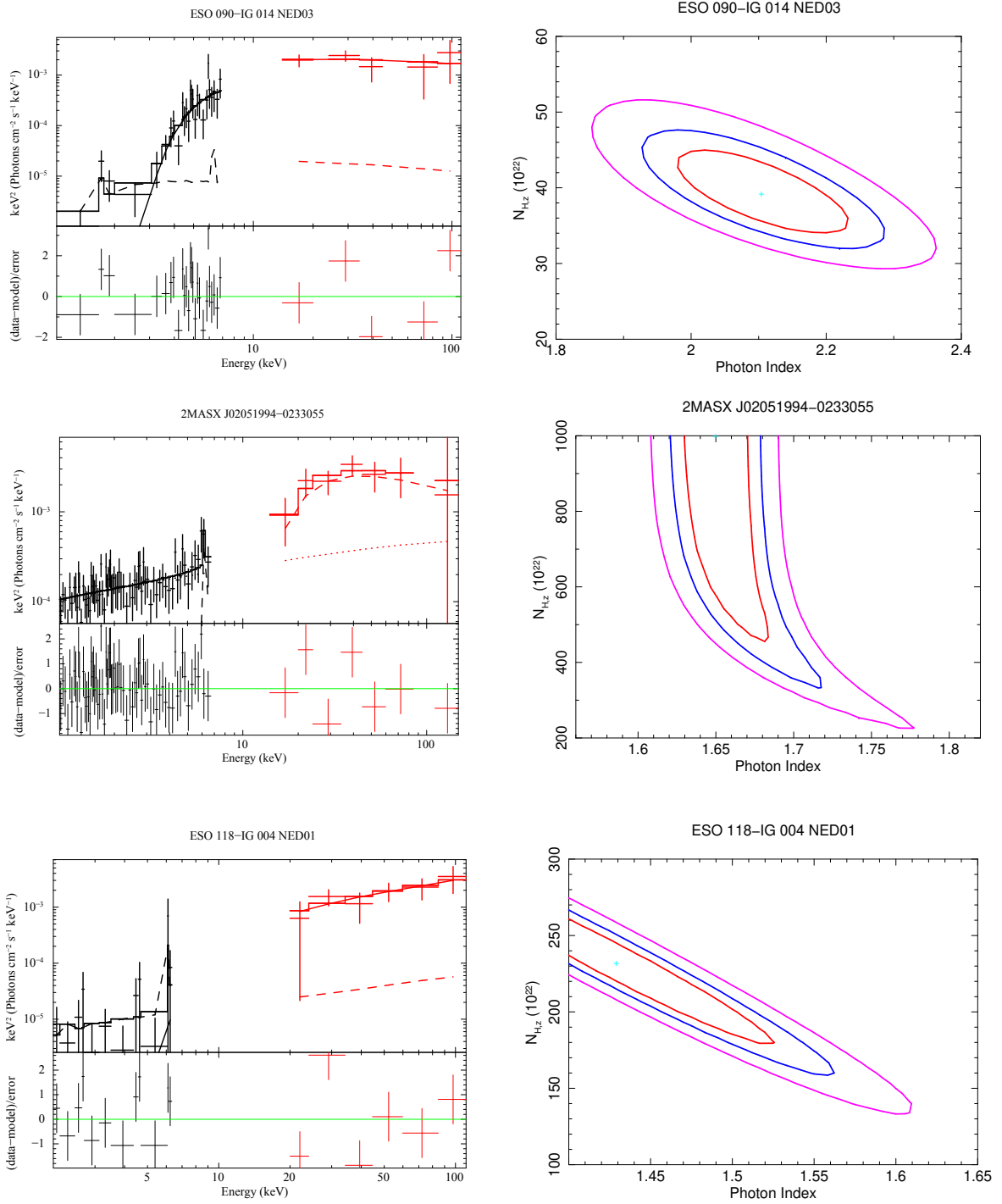}
    \caption{Unfolded \textit{Chandra} (black) and BAT (red) spectrum of each source fitted with the \bor model. The best-fitting line-of-sight component is plotted as a solid line, while the reflection component is a dashed line and the scattered component is a dotted line. The confidence contours at 68\%, 90\%, and 99\% are displayed for $\Gamma$ and N$_{H,Z}$ (in units of 10$^{22}$ cm$^{-2}$).}
    \label{fig:spec_cont1}
\end{figure*}

\begin{figure*}
    \centering
    \includegraphics[scale=0.8]{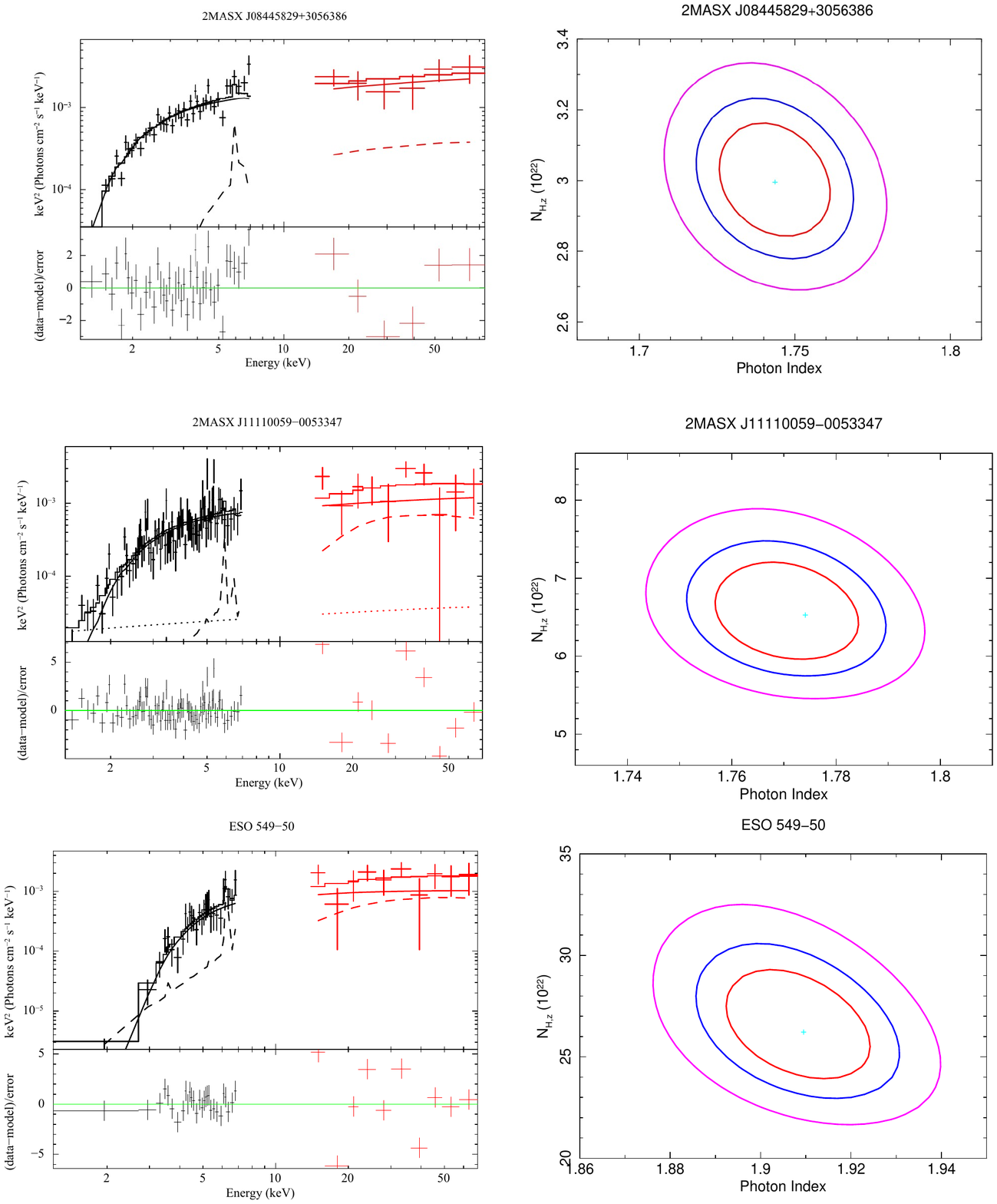}
    \caption{Unfolded \textit{Chandra} (black) and BAT (red) spectrum of each source fitted with the \bor model. The best-fitting line-of-sight component is plotted as a solid line, while the reflection component is a dashed line and the scattered component is a dotted line. The confidence contours at 68\%, 90\%, and 99\% are displayed for $\Gamma$ and N$_{H,Z}$ (in units of 10$^{22}$ cm$^{-2}$).}
    \label{fig:spec_cont2}
\end{figure*}

\begin{figure*}
    \centering
    \includegraphics[scale=0.8]{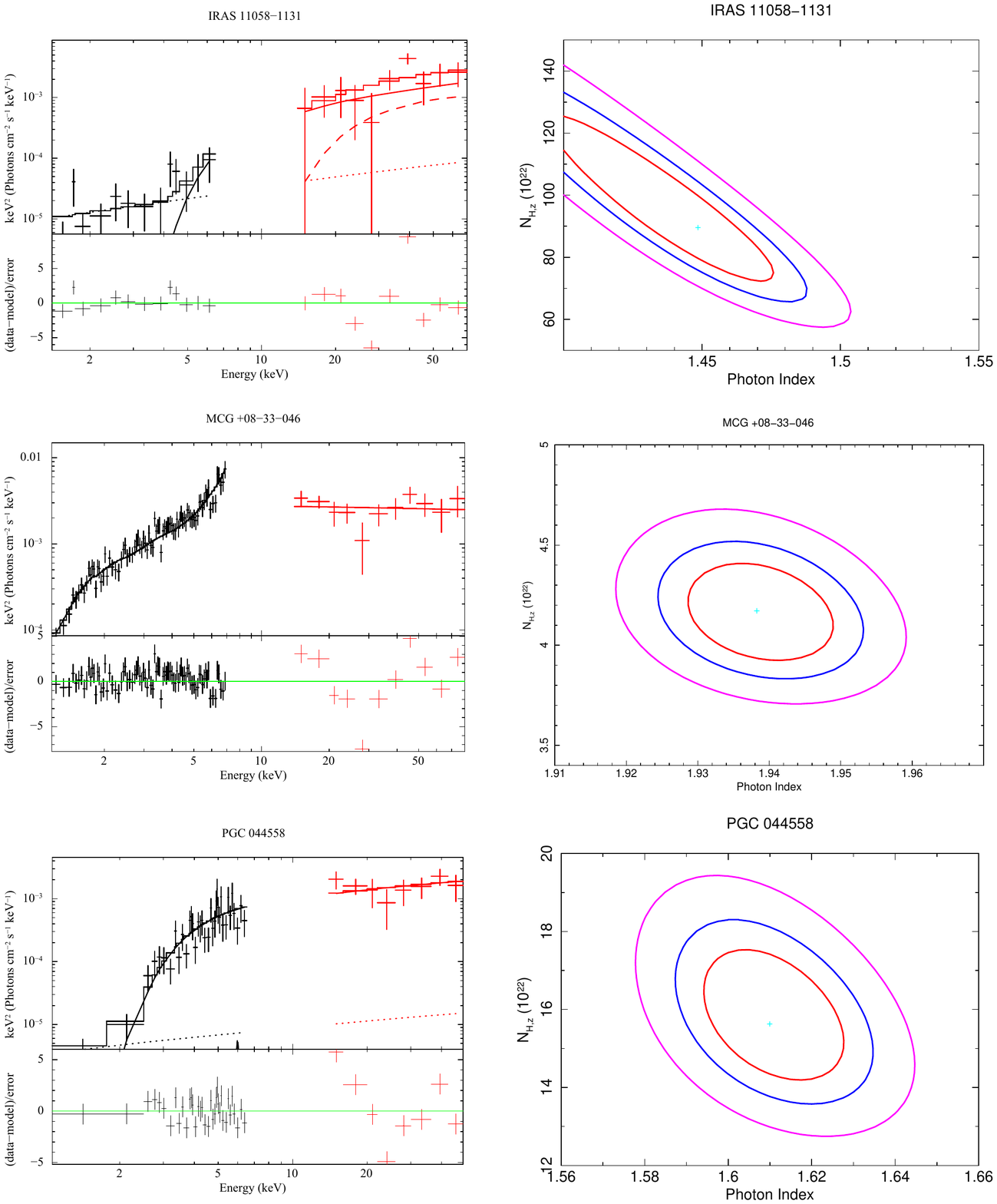}
    \caption{Unfolded \textit{Chandra} (black) and BAT (red) spectrum of each source fitted with the \bor model. The best-fitting line-of-sight component is plotted as a solid line, while the reflection component is a dashed line and the scattered component is a dotted line. The confidence contours at 68\%, 90\%, and 99\% are displayed for $\Gamma$ and N$_{H,Z}$ (in units of 10$^{22}$ cm$^{-2}$).}
    \label{fig:spec_cont3}
\end{figure*}

\begin{deluxetable*}{ccccccc} [h!]

\tablecaption{2MASX J02051994$-$0233055}
\label{tab:2masx02}

\tablehead{\colhead{Model} & \colhead{MYTorus} & \colhead{MYTorus} & \colhead{MYTorus} & \colhead{MYTorus} & \colhead{borus02} & \colhead{pexmon} \\ 
\colhead{} & \colhead{(Coupled)} & \colhead{(Decoupled Face-on)} & \colhead{(Decoupled Edge-on)} & \colhead{Face-on + Edge-on} & \colhead{} & \colhead{} }

\startdata
cstat/dof & 59/73 & 59/73 & 69/73 & 59/74 & 57/72 & 72/74 \\
$\Gamma$ & 1.78$^{+0.13}_{-0.10}$ & 1.73$^{+0.03}_{-0.13}$ & 1.76$^{+0.13}_{-0.14}$ & 1.76$^{+0.02}_{-0.03}$ & 1.65$^{+0.09}_{-0.22}$ & 1.83$^{+0.20}_{-0.19}$ \\
$N_{H,eq}$ & 10.00$^{-}_{-3.80}$ & ... & ... & ... & ... &  5.54$^{+5.76}_{-3.44}$\\
norm $10^{-2}$ & 0.64$^{+0.07}_{-0.06}$ & 0.18$^{+0.18}_{-0.02}$ & 0.13$^{+0.13}_{-0.07}$ & 0.18$^{+0.02}_{-0.01}$ & 0.27$^{+1.23}_{-0.16}$ & 0.09$^{+0.10}_{-0.05}$ \\
c$_{f,Tor}$ & ... & ... & ... & ... & 0.87$^{+0.11}_{-0.46}$  & ...\\
cos($\theta_{obs}$) & 0.25$^{+0.08}_{-0.08}$ & ... & ... & ... & 0.78$^{+0.21}_{-0.51}$  & 0.5*\\
$N_{H, Z}$ & ... & 10.00$^{-}_{-7.40}$  & 0.86$^{+0.40}_{-0.25}$ & 10.00$^{-}_{-7.30}$ &  10.00* & ... \\
$N_{H, S}$ & ... &  10.00$^{-}_{-5.80}$ & 5.0$^{-}_{-4.05}$ & 10.00$^{-}_{-5.80}$ & 10.00$^{-}_{-6.45}$ & ... \\
$f_s$ 10$^{-2}$ & 1.90$^{+4.60}_{-1.29}$ & 6.69$^{+5.31}_{-3.09}$ & 9.29$^{+8.01}_{-4.04}$ & 6.72$^{+5.28}_{-3.22}$ & 4.08$^{+11.90}_{-2.63}$ &  13.19$^{+16.49}_{-7.71}$ \\
C$_{cha}$ & 1* & 1* & 1* & 1* & 1* & 1* \\
\hline
L$\rm_{2-10\,keV}$ & & & & & 2.08$^{+0.08}_{-0.10}$ $\times$ 10$^{43}$\\
L$\rm_{15-55\,keV}$ & & & & & 3.00$^{+0.70}_{-0.59}$ $\times$ 10$^{43}$ \\
\enddata
\vspace{5mm}
\textbf{Notes:} Same as Table \ref{tab:eso09}. \\ 

\end{deluxetable*}

\begin{deluxetable*}{ccccccc}[h!]

\tablecaption{2MASX J04075215$-$6116126}
\label{tab:eso118}

\tablehead{\colhead{Model} & \colhead{MYTorus} & \colhead{MYTorus} & \colhead{MYTorus} & \colhead{borus02} \\ 
\colhead{} & \colhead{(Coupled)} & \colhead{(Decoupled Face-on)} & \colhead{(Decoupled Edge-on)} & \colhead{}}

\startdata
cstat/dof & 57/32 & 56/32 & 57/32 & 57/31 \\
$\Gamma$ & 1.61$^{+0.12}_{-0.12}$ & 1.60$^{+0.13}_{-0.10}$ & 1.60$^{+0.13}_{-0.11}$ & 1.56$^{+0.09}_{-0.14}$ \\
$N_{H,eq}$ & 0.30$^{+0.59}_{-0.15}$ & ... & ... & ... \\
norm $10^{-2}$ & 0.04$^{+0.03}_{-0.02}$ & 0.04$^{+0.03}_{-0.02}$ & 0.04$^{+0.03}_{-0.02}$ & 0.04$^{+0.02}_{-0.01}$ \\
c$_{f,Tor}$ & ... & ... & ... & 0.90$^{-}_{-0.85}$ \\
cos($\theta_{obs}$) & 0.37$^{+0.12}_{-}$ & ... & ... & 0.90$^{-}_{-}$ \\
$N_{H, Z}$ & ... & 0.20$^{+0.07}_{-0.06}$ & 0.20$^{+0.08}_{-0.06}$ & 0.21$^{+0.04}_{-0.08}$ \\
$N_{H, S}$ & ... & 0.39$^{+0.57}_{-}$ & 0.20$^{+0.64}_{-}$ & 0.19$^{+2.32}_{-0.15}$ \\
$f_s$ 10$^{-2}$ & 2.37$^{+3.10}_{-}$ & 2.37$^{+2.90}_{-2.30}$ & 2.54$^{+3.10}_{-}$ & 2.03* \\
C$_{cha}$ & 1* & 1* & 1* & 1* \\
\hline
L$\rm_{2-10\,keV}$ & & & & 2.09$^{+0.37}_{-0.35}$ $\times$ 10$^{42}$\\
L$\rm_{15-55\,keV}$ & & & & 3.16$^{+0.31}_{-0.28}$ $\times$ 10$^{42}$ \\
\enddata
\vspace{5mm}
\textbf{Notes:} Same as Table \ref{tab:eso09}. \\ 
\end{deluxetable*}

\begin{deluxetable*}{cccccc}[h!]

\tablecaption{2MASX J08445829$+$3056386} 
\label{tab:2masx08}

\tablehead{\colhead{Model} & \colhead{MYTorus} & \colhead{MYTorus} & \colhead{MYTorus} & \colhead{borus02} \\ 
\colhead{} & \colhead{(Coupled)} & \colhead{(Decoupled Face-on)} & \colhead{(Decoupled Edge-on)} & \colhead{} }

\startdata
cstat/dof & 204/171 & 203/171 & 205/171 & 201/170 \\
$\Gamma$ & 1.77$^{+0.05}_{-0.04}$ & 1.77$^{+0.07}_{-0.05}$ & 1.75$^{+0.06}_{-0.05}$ & 1.74$^{+0.07}_{-0.06}$  \\
$N_{H,eq}$ & 0.43$^{+0.44}_{-0.30}$ & ... & ... & ... \\
norm $10^{-2}$ & 0.10$^{+0.01}_{-0.01}$ & 0.10$^{+0.01}_{-0.01}$ & 0.10$^{+0.01}_{-0.01}$ & 0.09$^{+0.01}_{-0.01}$  \\
c$_{f,Tor}$ & ... & ... & ... & 1.00$^{-}_{-0.65}$  \\
cos($\theta_{obs}$) & 0.50$^{-}_{-0.02}$ & ... & ... & 0.05$^*$  \\
$N_{H, Z}$ & ... & 0.031$^{+0.003}_{-0.003}$ & 0.031$^{+0.003}_{-0.003}$ & 0.030 $^{+0.003}_{-0.003}$ \\
$N_{H, S}$ & ... & 0.50$^{+0.62}_{-0.35}$ & 0.30$^{+0.42}_{-0.23}$ & 0.24$^{+0.22}_{-0.15}$ \\
$f_s$ 10$^{-2}$ & 0.02$^{+2.18}_{-}$ & 0.01$^{+2.29}_{-}$ & 0.01$^{+2.34}_{-}$ & 0* \\
C$_{cha}$ & 1* & 1* & 1* & 1* \\
\hline
L$\rm_{2-10\,keV}$ & & & & 3.31$^{+0.16}_{-0.15}$ $\times$ 10$^{43}$\\
L$\rm_{15-55\,keV}$ & & & & 3.89$^{+0.38}_{-0.34}$ $\times$ 10$^{43}$ \\
\enddata
\vspace{5mm}
\textbf{Notes:} Same as Table \ref{tab:eso09}. \\ 
\end{deluxetable*}

\begin{deluxetable*}{ccccc}[h!]

\tablecaption{2MASX J11110059$-$0053347} 
\label{tab:2m111}

\tablehead{\colhead{Model} & \colhead{MYTorus} & \colhead{MYTorus} & \colhead{MYTorus} & \colhead{borus02} \\ 
\colhead{} & \colhead{(Coupled)} & \colhead{(Decoupled Face-on)} & \colhead{(Decoupled Edge-on)} & \colhead{}}
\startdata
cstat/dof & 98/74 & 99/74 & 97/74 & 98/74 \\
$\Gamma$ & 1.63$^{+0.07}_{-0.06}$ & 1.64$^{+0.07}_{-0.08}$ & 1.69$^{+0.09}_{-0.08}$ & 1.77$^{+0.02}_{-0.19}$ \\
$N_{H,eq}$ & 0.07$^{+0.25}_{-0.01}$ & ... & ... & ... \\
norm $10^{-2}$ & 0.05$^{+0.01}_{-0.01}$ & 0.06$^{+0.01}_{-0.02}$ & 0.06$^{+0.02}_{-0.01}$ & 0.06$^{+0.02}_{-0.01}$ \\
c$_{f,Tor}$ & ... & ... & ... & 0.89$^{-}_{-}$ \\
cos($\theta_{obs}$) & 0.0$^{+0.5}_{-}$ & ... & ... & 0.89$^{-}_{-}$ \\
$N_{H, Z}$ & ... & 0.07$^{+0.02}_{-0.02}$ & 0.06$^{+0.02}_{-0.02}$ & 0.07$^{+0.02}_{-0.02}$ \\
$N_{H, S}$ & ... & 0.01$^{-}_{-}$ & 0.96$^{-}_{-}$ & 31.62$^{-}_{-}$ \\ 
$f_s$ & 0.04$^{+0.04}_{-}$ & 0.04$^{+0.04}_{-}$ & 0.03$^{+0.04}_{-}$ & 0.03$^{+0.04}_{-0.03}$ \\
C$_{cha}$ & 1* & 1* & 1* & 1* \\
\hline
L$\rm_{2-10\,keV}$ & & & & 3.98$^{+0.38}_{-0.35}$ $\times$ 10$^{43}$\\
L$\rm_{15-55\,keV}$ & & & & 5.01$^{+0.48}_{-0.44}$ $\times$ 10$^{43}$ \\
\enddata
\vspace{0.5cm}
\textbf{Notes:} Same as Table \ref{tab:eso09}. \\ 
cstat/dof: \textit{Chandra} data only.
\end{deluxetable*}

\begin{deluxetable*}{ccccc}[h!]

\tablecaption{ESO 549$-$50}
\label{tab:es0549}

\tablehead{\colhead{Model} & \colhead{MYTorus} & \colhead{MYTorus} & \colhead{MYTorus} & \colhead{borus02} \\ 
\colhead{} & \colhead{(Coupled)} & \colhead{(Decoupled Face-on)} & \colhead{(Decoupled Edge-on)} & \colhead{}}
\startdata
cstat/dof & 22/13 & 21/13 & 22/23 & 21/23 \\
$\Gamma$ & 1.81$^{+0.25}_{-0.14}$ & 1.86$^{+0.23}_{-0.15}$ & 1.81$^{+0.18}_{-0.10}$ & 1.91$^{+0.02}_{-0.02}$ \\
$N_{H,eq}$ & 0.27$^{-}_{-0.05}$ & ... & ... & ... \\
norm $10^{-2}$ & 0.10$^{+0.09}_{-0.04}$ & 0.10$^{+0.10}_{-0.04}$ & 0.10$^{+0.09}_{-0.03}$ & 0.10$^{+0.10}_{-0.02}$ \\
c$_{f,Tor}$ & ... & ... & ... & 0.35$^{+0.05}_{-0.05}$ \\
cos($\theta_{obs}$) & 0.01$^{+0.49}_{-}$ & ... & ... & 0.95$^{-}_{-0.77}$ \\
$N_{H, Z}$ & ... & 0.25$^{+0.07}_{-0.04}$ & 0.26$^{+0.14}_{-0.06}$ & 0.26$^{+0.08}_{-0.06}$ \\
$N_{H, S}$ & ... & 1.6$^{-}_{-1.59}$ & 0.50$^{-}_{-}$ & 14.13$^{-}_{-}$ \\ 
$f_s$ & 0.002$^{+0.004}_{-}$ & 0.001* & 0.002* & 0 \\
C$_{cha}$ & 1* & 1* & 1* & 1* \\
\hline
L$\rm_{2-10\,keV}$ & & & & 3.98$^{+0.70}_{-0.59}$ $\times$ 10$^{42}$\\
L$\rm_{15-55\,keV}$ & & & & 3.72$^{+0.36}_{-0.33}$ $\times$ 10$^{42}$ \\
\enddata
\vspace{0.5cm}
\textbf{Notes:} Same as Table \ref{tab:2m111}. \\ 
\end{deluxetable*}

\begin{deluxetable*}{ccccc}[h!]

\tablecaption{IRAS 11058$-$1131}
\label{tab:iras}

\tablehead{\colhead{Model} & \colhead{MYTorus} & \colhead{MYTorus} & \colhead{MYTorus} & \colhead{borus02} \\ 
\colhead{} & \colhead{(Coupled)} & \colhead{(Decoupled Face-on)} & \colhead{(Decoupled Edge-on)} & \colhead{}}
\startdata
cstat/dof & 11/6 & 11/6 & 11/6 & 11/6 \\
$\Gamma$ & 1.41$^{+0.61}_{-}$ & 1.45$^{+0.25}_{-}$ & 1.41$^{+0.40}_{-}$ & 1.45$^{+0.35}_{-}$ \\
$N_{H,eq}$ & 10.00$^{-}_{-}$ & ... & ... & ... \\
norm $10^{-2}$ & 0.20$^{+3.30}_{-0.20}$ & 0.04$^{+0.04}_{-0.03}$ & 0.04$^{+0.12}_{-0.02}$ & 0.04$^{+0.16}_{-0.02}$ \\
c$_{f,Tor}$ & ... & ... & ... & 0.60$^{-}_{-}$ \\
cos($\theta_{obs}$) & 0.15$^{+0.35}_{-}$ & ... & ... & 0.05* \\
$N_{H, Z}$ & ... & 2.60$^{-}_{-1.93}$ & 0.94$^{-}_{-0.93}$ & 0.81$^{+0.84}_{-0.74}$ \\
$N_{H, S}$ & ... & 9.04$^{-}_{-6.88}$ & 8.75$^{-}_{-}$ & 7.41$^{-}_{-}$ \\ 
$f_s$ & 0.003$^{+0.099}_{-0.003}$ & 0.02$^{+0.03}_{-0.01}$ & 0.02$^{-}_{-0.02}$ & 0.03$^{-}_{-}$ \\
C$_{cha}$ & 1.13$^{+5.27}_{-1.02}$ & 0.88$^{+0.91}_{-0.87}$ & 0.92$^{-}_{-0.91}$ & 0.70$^{+5.30}_{-0.69}$ \\
\hline
L$\rm_{2-10\,keV}$ & & & & 1.55$^{+1.47}_{-1.06}$ $\times$ 10$^{43}$\\
L$\rm_{15-55\,keV}$ & & & & 3.16$^{+0.31}_{-0.28}$ $\times$ 10$^{43}$ \\
\enddata
\vspace{0.5cm}
\textbf{Notes:} Same as Table \ref{tab:2m111}. \\ 
BAT cross-norm fixed at 1.
 
\end{deluxetable*}

\begin{deluxetable*}{cccccc}[h!]

\tablecaption{MCG $+$08-33-046}
\label{tab:mcg}

\tablehead{\colhead{Model} & \colhead{pileup} \\ 
\colhead{} & \colhead{}}
\startdata
cstat/dof & 86/77 \\
$\Gamma$ & 2.05\\
$N_{H,eq}$ & 0.02 \\
norm $10^{-2}$ & 0.32\\
c$_{f,Tor}$ &  ...\\
cos($\theta_{obs}$) & ... \\
$N_{H, Z}$ & ... \\
$N_{H, S}$ & ... \\ 
$f_s$ & ... \\
C$_{cha}$ & 1* \\
\hline
L$\rm_{2-10\,keV}$ & 4.90$^{+0.47}_{-0.43}$ $\times$ 10$^{42}$\\
L$\rm_{15-55\,keV}$ & 3.55$^{+0.17}_{-0.16}$ $\times$ 10$^{42}$ \\
\enddata
\vspace{0.5cm}
\textbf{Notes:} Same as Table \ref{tab:2m111}. \\ 
\end{deluxetable*}

\begin{deluxetable*}{ccccc}[h!]

\tablecaption{IRAS 12571$+$7643}
\label{tab:pgc}

\tablehead{\colhead{Model} & \colhead{MYTorus} & \colhead{MYTorus} & \colhead{MYTorus} & \colhead{borus02} \\ 
\colhead{} & \colhead{(Coupled)} & \colhead{(Decoupled Face-on)} & \colhead{(Decoupled Edge-on)} & \colhead{}}
\startdata
cstat/dof & 44/30 & 44/30 & 43/30 & 41/30 \\
$\Gamma$ & 1.69$^{+0.11}_{-0.11}$ & 1.71$^{+0.12}_{-0.10}$ & 1.75$^{+0.13}_{-0.14}$ & 1.61$^{+0.18}_{-0.07}$ \\
$N_{H,eq}$ & 0.17$^{+0.45}_{-0.03}$ & ... & ... & ... \\
norm $10^{-2}$ & 0.07$^{+0.02}_{-0.01}$ & 0.08$^{+0.01}_{-0.01}$ & 0.07$^{+0.03}_{-0.02}$ & 0.05$^{+0.02}_{-0.01}$ \\
c$_{f,Tor}$ & ... & ... & ... & 0.15$^{-}_{-}$ \\
cos($\theta_{obs}$) & 0.0$^{+0.5}_{-}$ & ... & ... & 0.90$^{-}_{-}$ \\
$N_{H, Z}$ & ... & 0.18$^{+0.06}_{-0.03}$ & 0.17$^{+0.05}_{-0.05}$ & 0.16$^{+0.05}_{-0.03}$ \\
$N_{H, S}$ & ... & 0.02$^{+0.44}_{-}$ & 0.90$^{-}_{-}$ & 0.01$^{-}_{-}$ \\ 
$f_s$ & 0.01$^{+0.01}_{-0.01}$ & 0.01$^{+0.01}_{-0.01}$ & 0.01$^{+0.02}_{-0.01}$ & 0.01$^{+0.01}_{-}$ \\
C$_{cha}$ & 1* & 1* & 1* & 1* \\
\hline
L$\rm_{2-10\,keV}$ & & & & 2.09$^{+0.25}_{-0.27}$ $\times$ 10$^{43}$\\
L$\rm_{15-55\,keV}$ & & & & 3.63$^{+0.17}_{-0.16}$ $\times$ 10$^{43}$ \\
\enddata
\vspace{0.5cm}
\textbf{Notes:} Same as Table \ref{tab:2m111}. \\ 
 
\end{deluxetable*}

\section{\textit{Chandra} Exposures}\label{AppB}

\begin{figure*}
    \centering
    \includegraphics[scale=0.8]{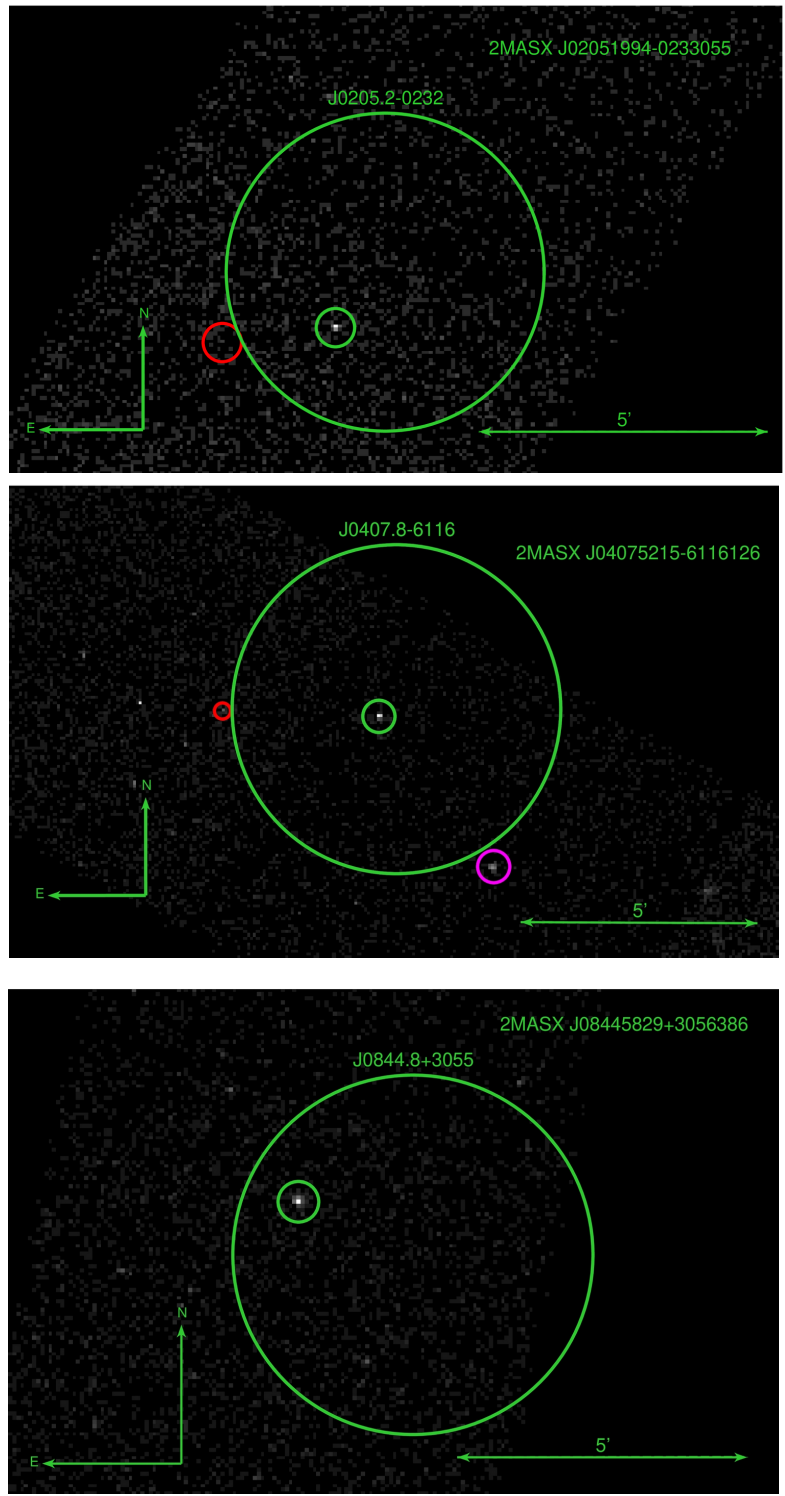}
    \caption{From top to bottom: the \textit{Chandra} exposures for 2MASX J02051994$-$0233055, 2MASX J04075215$-$6116126, 2MASX J08445829$+$3056386 \textbf{in the 1$-$7\,keV range}. Large green circles represent the BAT 95\,\% uncertainty region while small green circles identify the correct counterpart of the BAT source; red/magenta circles are for sources detected by \textit{Chandra} and located within/nearby the BAT 95\,\% uncertainty region, but which we determine are not the counterparts of the BAT source (see the text for more details).}
    \label{fig:field1}
\end{figure*}

\begin{deluxetable}{ccc}
\tablecaption{BAT 95\% Confidence Regions} 
\label{tab:batregions}

\tablehead{\colhead{\swift-BAT ID} &  \colhead{Source Name} & \colhead{95\% Region} \\
\colhead{} & \colhead{} & \colhead{Arcmin}}
\startdata
J0205.2$-$0232 & 2MASX J02051994$-$0233055$^*$ & 2.76 \\
J0402.6$-$2107 & ESO 549$-$50$^{\dagger}$ & 4.49 \\
J0407.8$-$6116 & 2MASX J04075215$-$6116126$^*$ & 3.40 \\
J0844.8$+$3055 & 2MASX J08445829$+$3056386$^*$ & 2.95 \\
J0901.8$-$6418 & ESO 090$-$IG 014$^*$ & 2.87 \\
J1108.4$-$1148 & IRAS 11058$-$1131$^{\dagger}$ & 4.73 \\
J1111.0$+$0054 & 2MASX J11110059$-$0053347$^{\dagger}$ & 4.65 \\
J1258.4$+$7624 & IRAS 12571+7643$^{\dagger}$ & 4.35 \\
J1828.8$+$5021 & MCG $+$08-33-046$^{\dagger}$ & 3.77 
\enddata
Notes: \\
$*$: From the 100-month BAT Catalog. \\
$\dagger$: From the 150-month BAT Catalog. \\
\end{deluxetable}

\newpage
\begin{figure*}
    \centering
    \includegraphics[scale=0.8]{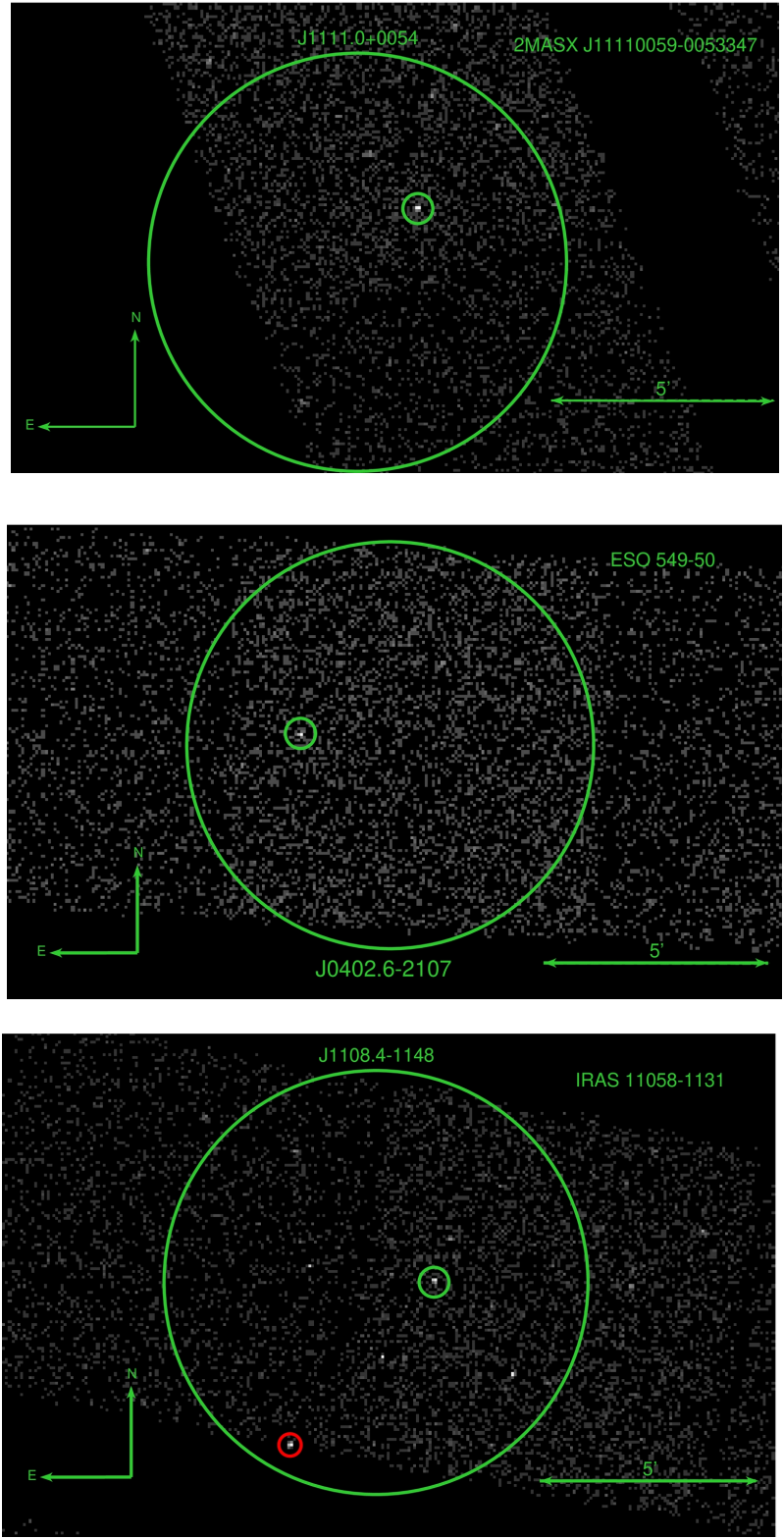}
    \caption{From top to bottom: the \textit{Chandra} exposures for 2MASX J11110059$-$0053347, ESO 549$-$50, IRAS 11058$-$1131 \textbf{in the 1$-$7\,keV range}. Large green circles represent the BAT 95\,\% uncertainty region while small green circles identify the correct counterpart of the BAT source; red/magenta circles are for sources detected by \textit{Chandra} and located within/nearby the BAT 95\,\% uncertainty region, but which we determine are not the counterparts of the BAT source (see the text for more details).}
    \label{fig:field2}
\end{figure*}

\begin{figure*}
    \centering
    \includegraphics[scale=0.8]{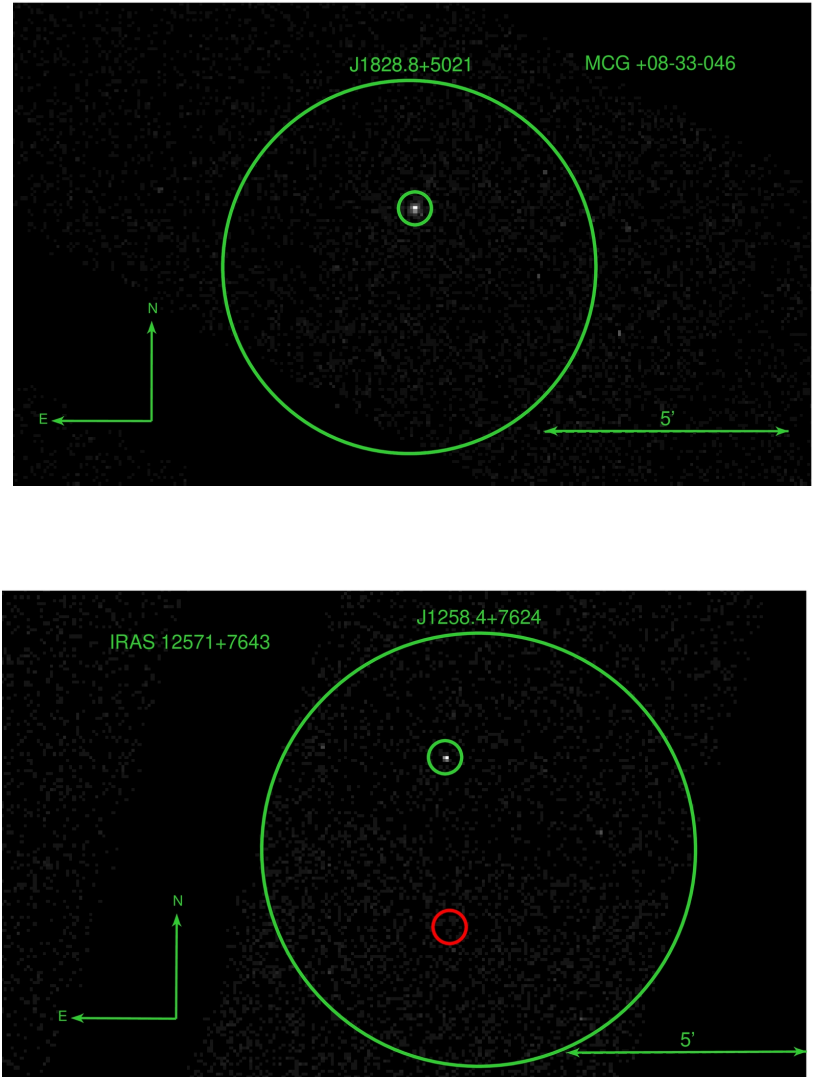}
    \vspace{-0.4cm}
    \caption{From top to bottom: the \textit{Chandra} exposures for MCG $+$08-33-046, IRAS 12571$+$7643 \textbf{in the 1$-$7\,keV range}. Large green circles represent the BAT 95\,\% uncertainty region while small green circles identify the correct counterpart of the BAT source; red/magenta circles are for sources detected by \textit{Chandra} and located within/nearby the BAT 95\,\% uncertainty region, but which we determine are not the counterparts of the BAT source (see the text for more details).}
    \label{fig:field3}
\end{figure*}

\bibliographystyle{apj}
\bibliography{bibliography}

\end{document}